\documentclass[
hidelinks,
twocolumn,
secnumarabic,
amssymb, 
nobibnotes, 
aps, 
prb]{revtex4-2}
\usepackage{graphicx}
\usepackage{multirow}
\usepackage{lipsum} 
\usepackage[breaklinks=true]{hyperref} % break urls in references
\usepackage[T1]{fontenc}
\usepackage[utf8]{inputenc}
\usepackage{textcomp} % for \textdegree
\usepackage{lmodern}
\graphicspath{{./figures/}}

\usepackage{comment}

\usepackage[colorinlistoftodos]{todonotes}
	\newcommand{\justyna}[1]{\todo[author=Justyna,size=\small,color=red!30]{#1}}
	\newcommand{\justynainline}[1]{\todo[author=Justyna,size=\small,inline,color=red!30]{#1}}

\usepackage{xcolor}
\hypersetup{
    breaklinks = true,
    colorlinks,
    pdftitle = {}, 
    pdfauthor = {},
    pdfkeywords = {},
    linkcolor={blue!80!black},
    citecolor={blue!80!black},
    urlcolor={blue!80!black},
    filecolor = black
}
\graphicspath{{./figures/}}

\begin{document}
\begin{sloppypar} %for overfull lines with long chemical formulas

\title{
Magnetic hardness of hexagonal and orthorhombic Fe$_{3}$C, Co$_{3}$C, (Fe-Co)$_{3}$C, and \\their alloys with boron, nitrogen, and transition metals: 
A first-principles study
}

\author{Justyn Snarski-Adamski}
\author{Miros\l{}aw Werwi\'nski}
\email[Corresponding author: Mirosław Werwiński\\Email address: ]{werwinski@ifmpan.poznan.pl}
\author{Justyna Rychły-Gruszecka}
\affiliation{Institute of Molecular Physics, Polish Academy of Sciences,  M. Smoluchowskiego 17, 60-179 Pozna\'n, Poland}
\date{\today}%

\begin{abstract}
%
%----intro---------
%
Restrictions on the availability of rare earth metals create a strong demand for new rare-earth-free hard magnetic materials.
In this study, we considered a large set of materials that are closely related to orthorhombic Fe$_3$C (cementite) with the aim of characterizing trends in their intrinsic magnetic properties, highlighting the relation between magnetic properties and the chemical composition, and identifying alloys that are optimal for applications.
%
%
%---------subjects of the investigation----------
%
A comprehensive analysis was conducted on the full concentration ranges of hexagonal ($\epsilon$) and orthorhombic ($\theta$)
phases of 
(Fe-Co)$_3$C, 
(Fe-Co)$_3$(B-C), 
(Fe-Co)$_3$(C-N), and 
their alloys with 3$d$, 4$d$ and 5$d$ transition metals. 
%
%----methods-------
%
The calculations were performed using the density functional theory implemented in the full-potential local-orbital code (FPLO).
%
%----results and conclusions--------
%
Calculated properties included 
formation energies, 
Curie temperatures, 
magnetic moments, 
magnetocrystalline anisotropy energies (MAE), and
magnetic hardnesses.
The considered compositions exhibit a range of magnetic properties, including soft, semi-hard, and hard magnetic.
The materials most promising for hard-magnetic applications are selected Co-rich orthorhombic and hexagonal (Fe,Co)$_3$C alloys.
The calculation results do not indicate that substituting with transition metals increases the potential of the alloys for permanent magnet applications.
A significant drawback of alloying orthorhombic $\theta$-Fe$_3$C (cementite) with transition metals is the notable decline in the Curie temperature.
Among the positive outcomes, we found that a considerable proportion of the orthorhombic Co$_3$(B-C-N) alloys are magnetically hard, of which boron substitution raises the Curie temperature and improves stability.
By mapping the dependence of MAE on the concentration of elements covering both the 3$d$ (from Fe to Co) and 2$p$ (from B, through C, to N) positions, we have demonstrated for the first time the near isoelectronic nature of MAE.
The latter observation may be particularly useful in designing compositions of new magnetically hard materials.
\end{abstract}

%\date{\today}

\maketitle

\section{Introduction}

%---------rare-earth free permanent magnets - motivations ----------------
%
Permanent magnets are an essential component of modern technology~\cite{gutfleisch_magnetic_2011, ronning_rare_2014, lewis_perspectives_2013}.
Demand for neodymium magnets reached 119.2 thousand tons in 2020 and is growing~\cite{smith_rare_2022}.
However, there are potential risks associated with the use of neodymium magnets, such as limited availability of neodymium and strong fluctuations of its price.
The significant increase in price of rare-earth metals that took place around 2011 is referred to as the \textit{2011 rare-earth crisis}~\cite{bourzac2011rare}. 
A similar economic situation was repeated in 2022.
For the above reasons, it is important to identify new materials for permanent magnet applications that require minimal or no rare earth elements~\cite{
kuzmin_towards_2014,
skokov_heavy_2018, 
cui_current_2018, 
li_prospect_2019, 
vishina_high-throughput_2020,
coey_perspective_2020}.

In this work, we will focus on Fe$_{3}$C-based alloys intensively studied for more than a decade as promising materials for permanent magnets~\cite{
zhang2011controlled, 
carroll_magnetic_2012, 
rohith_vinod_large_2015,
el2015experimental,
pal_properties_2017, 
wu2017exploring,
mohapatra_hard_2020,
marshall_high-pressure_2021}. 
%
%Moreover, Fe$_3$C nanoparticles are being studied experimentally for hyperthermia applications~\cite{wang_fe3c_2015, gangwar_fe3c_2019, gangwar_magnetic_2021}.

%-----materials considered in this work----------
%
In this study, we present a first-principles analysis of the hexagonal iron carbide $\epsilon$-Fe$_{3}$C (referred to hereafter as hexa-Fe$_{3}$C), the orthorhombic cementite $\theta$-Fe$_{3}$C (referred to hereafter as ortho-Fe$_{3}$C), and the corresponding Co$_{3}$C phases (hexa-Co$_{3}$C and ortho-Co$_{3}$C).
Moreover, we calculate a series of Fe$_{3}$C and  Co$_{3}$C alloys with transition metal elements 3$d$, 4$d$, and 5$d$ and also consider the (Fe,Co)$_{3}$C and (Fe,Co)$_{3}$(B,C,N) alloys.

%------experimental properties---------------
%
When discussing Fe$_3$C and Co$_3$C alloys for use as permanent magnets, we will focus primarily on two intrinsic properties of the materials, magnetocrystalline anisotropy energy (MAE) and Curie temperature ($T_C$).
%
%--------Fe3C-------
%
The experimental value of the uniaxial magnetocrystalline anisotropy constant of ortho-Fe$_{3}$C is a moderate 0.405~MJ\,m$^{-3}$ at 5~K~\cite{choe_easy_2016}.
%and the easy axis of magnetization is the $c$-axis.
%hexagonal Fe$_{3}$C is about 0.25~MJ\,m$^{-3}$ - no source
%
Whereas, the measured Curie temperature for ortho-Fe$_{3}$C has been measured as 481~K~\cite{nicholson_solubility_1957} and 483~K~\cite{tsuzuki_high_1984}.
The state of the knowledge on the properties of ortho-Fe$_{3}$C (cementite) is summarized by Bhadeshia~\cite{bhadeshia2020cementite}. 
%
%---------Co3C------
%
For ortho-Co$_{3}$C, the measured  Curie temperatures are 498~K~\cite{mikhalev_magnetic_2019}, 
510~K~\cite{harris_high_2010}, 
563~K~\cite{turgut_metastable_2016}, 
and 650~K~\cite{el-gendy_enhanced_2014}.
The results were determined for nanoparticles, and their large dispersion indicates a significant effect of structurization on the magnetic properties of the system.
The ortho-Co$_{3}$C shows an experimental MAE of 0.74$\pm$0.1~MJ\,m$^{-3}$~\cite{el-gendy_enhanced_2014}.
Additionally, Co$_{3}$C carbide nanocrystals have been confirmed to possess hard magnetic properties~\cite{zhang2011controlled}.
Enhanced magnetocrystalline anisotropy has also been observed in Co-C nanowires, Co-C nanoparticles, and polyphase Fe-Co-C polycrystalline samples~\cite{el2013coxc, el-gendy_enhanced_2014, el2015experimental}.
%
%The coercivity measured for Co$_{3}$C at room temperature is equal 3.1~kOe with energy product (BH)$_{max}$ of 20.7~kJ\,m$^{-3}$ and saturation magnetization M$_{S}$ of 72~emu\,g$^{-1}$~\cite{harris_high_2010}. 
%
%
%-------(Fe,Co)3C-------
%
The melt-spun Fe$_2$CoC alloys show elevated magnetocrystalline anisotropy constant ($K_1$) of 0.91~MJ\,m$^{-3}$ and magnetic moment of 1.23~$\mu_\mathrm{B}$\,atom$^{-1}$ (at 10~K)~\cite{wu2017exploring}.
In contrast, an exceptionally high $K_1$ of 4.6~MJ\,m$^{-3}$ has been measured for Fe$_2$CoC nanoparticles~\cite{el2015experimental}.

%-------Fe3C+TM  and Co3C+TM-------
%
%In addition to discussed (Fe,Co)$_{3}$C alloys, 
%
Fe$_3$C compositions with transition metals were also experimentally studied, among others: Ti, V, Cr, Mn, Ni, and Mo~\cite{
kagawa_lattice_1979, 
schaaf_mossbauer_1992,
umemoto_influence_2001,
wang_fe3c_2015}.
It was determined what effect the selected substitutions have on the lattice parameters~\cite{kagawa_lattice_1979} and Curie temperature~\cite{schaaf_mossbauer_1992}.
It was also found that Cr, Mn, V, and Mo stabilize the alloy, while Ti and Ni destabilize it~\cite{umemoto_influence_2001}.

%--------Fe3(B,C,N) and Co3(B,C,N)
%
Another type of modification of Fe$_3$C that significantly affects properties is alloying with boron or nitrogen in place of carbon~\cite{nicholson_solubility_1957, rounaghi_synthesis_2019}.
For example, it has been shown experimentally that going from Fe$_3$C to Fe$_3$B the Curie temperature increases linearly from 481~K to 824~K~\cite{nicholson_solubility_1957}.

\begin{table*}[t]
\center
\caption{
Total spin magnetic moments [total $m_{s}$ ($\mu_\mathrm{B}$\,f.u.$^{-1}$)] calculated for hexa-Fe$_{3}$C, hexa-Co$_{3}$C, ortho-Fe$_{3}$C and ortho-Co$_{3}$C, together with contributions from Fe, Co, and C sites [$m_{s}$(Fe/Co/C) ($\mu_\mathrm{B}$\,atom$^{-1}$)].
For orthorhombic phases, magnetic moments on general and special positions of Fe and Co are presented.
Our results obtained using the FPLO18 code with PBE exchange-correlation potential are compared with previous calculation outcomes from literature.
}
\label{tab:literature}
\begin{tabular}{lcccc}
        \hline
        \hline
        hexa-Fe$_{3}$C & & $m_{s}$(Fe) & $m_{s}$(C)  & total $m_{s}$ \\ %& MAE \\
        \hline
        FPLO, GGA(PBE) [this work] & & 2.12 & -0.30 & 6.04 \\ % & 0.68 \\
        %\hline
        VASP, PAW, GGA \cite{wu2017exploring} & & - & - & 5.92 \\
        CASTEP, PBE-GGA \cite{zhang2012electronic, hui_stability_2018} & & (2.04 -- 2.23) & (-0.26 -- -0.24) & (5.90 - 6.42) \\
        \hline
        hexa-Co$_{3}$C & & $m_{s}$(Co) & $m_{s}$(C)  & total $m_{s}$ \\ % & MAE \\
        \hline
        FPLO, GGA(PBE) [this work] & & 1.03 & -0.12 & 2.95 \\ %& -0.15 \\
        %\hline
        VASP, PAW, GGA \cite{wu2017exploring} & & - & - & 2.76 \\
        \hline
        ortho-Fe$_{3}$C & $m_{s}$(Fe$^{g}$) & $m_{s}$(Fe$^{s}$) & $m_{s}$(C)  & total $m_{s}$ \\ %& MAE \\
        \hline
        FPLO, GGA(PBE) [this work] & 1.93 & 2.01 & -0.28 & 5.56 \\ %& 0.11 \\
        VASP, PAW, GGA \cite{medvedeva_effect_2008, shein_electronic_2007, buggenhoudt_predicting_2021, fang_structural_2009, dick2011ab, shein_electronic_2006, wu2017exploring} & (1.84 - 1.93) & (1.92 - 1.98) & (-0.12 - -0.14) & (5.52 - 5.72) \\
        % %
        VASP, PW91, USPP-GGA \cite{chiou2003structure} & 1.95 & 1.99 & -0.16 & 5.73 \\    
        WIEN2K, FP-LAPW, GGA \cite{faraoun2006crystalline} & 1.96 & 1.97 & -0.13 & 5.76 \\
        QUANTUM ESPRESSO, PWscf, GGA \cite{odkhuu_substitution-_2016} & 2.04 & 2.09 & -0.24 & 5.90 \\
        CASTEP, GGA \cite{hui_stability_2018, chen2023first, wu2023atomistic} & (1.83 -1.90) & (1.94 - 1.98) & (-0.24 -- -0.23) & (5.55 - 5.56) \\
        \hline
        ortho-Co$_{3}$C & $m_{s}$(Co$^{g}$) & $m_{s}$(Co$^{s}$) & $m_{s}$(C)  & total $m_{s}$ \\ %& MAE \\
        \hline
        FPLO, GGA(PBE) [this work] & 1.12 & 1.02 & -0.11 & 3.16 \\ % & 0.65 \\
        %\hline
        VASP, PAW, GGA \cite{shein_electronic_2006, wu2017exploring} & 0.98 & 1.07 & -0.05 & (2.98 - 3.12) \\
        \hline
        \hline
    \end{tabular}
\end{table*}

%------DFT results---------------
%
Recently, based on first-principles calculations combined with a structure prediction algorithm, positive formation energies (indicating instability) have been determined for the well-known iron carbides and for a large set of newly determined iron carbides~\cite{yuan2020crystal}.
Direct studies on Fe$_{3}$C were also carried out using first-principle calculations~\cite{lv2008first, zhang2012electronic}. 
The average spin magnetic moments have been estimated as 1.50 and 1.40~$\mu_{B}$\,atom$^{-1}$ for the hexa- and ortho-Fe$_{3}$C phases, respectively~\cite{lv2008first}.
For details, like spin magnetic moment on Fe and C sites, please check the Table~\ref{tab:literature}.

%-------cementite with TM on Fe site - DFT results----------
%
To improve stability and determine magnetic properties, first-principles calculations have been repeatedly used to model cementite alloys with Cr~\cite{medvedeva_effect_2008, konyaeva_electronic_2009, lv2011first_cementiteCrMn, odkhuu_substitution-_2016,yang2016first, buggenhoudt_predicting_2021}, Mn~\cite{lv2011first_cementiteCrMn, buggenhoudt_predicting_2021}, 
Co~\cite{wang2011electronic, gao2014first, wu2017exploring}, and
Ni~\cite{wang2011electronic, gao2014first}.
Whereas, the broader context is complemented by calculations of wide ranges of transition metal substitutions in ortho-Fe$_{3}$C~\cite{shein_electronic_2007, ande2012first, razumovskiy2015first}.

%-------alloys substituted on C site - DFT----------
%
First-principles calculations were also made for orthorhombic cementite-type and hexagonal $\epsilon$ phases of Fe$_{3}$B and Fe$_{3}$N compounds~\cite{lv2010structural, zhang2012electronic}.
The full range of alloy concentrations of ortho-Fe$_{3}$(C$_{1-y}$B$_{y}$) was modeled by the supercell method~\cite{medvedeva2007simulation}.
Furthermore, Al, Si, P, and S were considered at the C site in the cementite~\cite{ande2012first}.
In addition, in our previous work we already considered Fe$_{0.7}$Co$_{0.3}$ thin films with B, C, and N atoms in interstitial positions~\cite{marciniak_first-principles_2024}.

%------section summary----------
%
This study, covers the wide ranges of alloying with transition metals, boron, nitrogen, and, going beyond previous reports, focuses on determination of magnetic hardnesses and Curie temperatures -- two parameters especially important for permanent magnet applications.

\section{\label{Calculation details}Calculations details}

\begin{table*}[t]
\center
\caption{
The initial and optimized lattice parameters and Wyckoff positions for hexa-Fe$_{3}$C (s.g. $P$6$_{3}$22) and ortho-Fe$_{3}$C (s.g. $Pnma$) phases.
The geometry optimization was performed with the FPLO18 code using the PBE potential.
}
\label{tab:struct}
\begin{tabular}{ c  c  c  c  c }
        \hline
        \hline
        \multicolumn{4}{c}{Initial structural data} \\
        \hline
        Phase & Space group & Lattice parameters & Wyckoff positions\\
        \hline
         %   & & & \begin{tabular}{c c c} x & y & z \end{tabular}\\
        %\hline
        hexa-Fe$_{3}$C \cite{fruchart1984etudes} 
            & $P$6$_{3}$22 (182)
            & 
                \begin{tabular}{c} 
                    $a$ = $b$ = 4.767 \\
                    $c$ = 4.354 \\
                \end{tabular}
              & 
                \begin{tabular}{c c c c} 
                    Fe = & 0.333 & 0 & 0 \\
                    C = & 1/3 & 2/3 & 3/4 \\
                \end{tabular} \\
        \hline
        ortho-Fe$_{3}$C \cite{yakel1985crystal} 
            & $Pnma$ (62)
            & 
                \begin{tabular}{c} 
                    $a$ = 5.092 \\
                    $b$ = 6.741 \\
                    $c$ = 4.527 \\
                \end{tabular}
            & 
                \begin{tabular}{c c c c} 
                    Fe$^{g}$ = & 0.1834 & 0.0689 & 0.3344 \\
                    Fe$^{s}$ = & 0.0388 & 0.25 & 0.8422 \\
                    C = & 0.8764 & 0.25 & 0.4426 \\
                \end{tabular} \\
        \hline
        \multicolumn{4}{c}{Optimized structural data} \\
        \hline
        Phase & Space group & Lattice parameters & Wyckoff positions\\
        \hline
        hexa-Fe$_{3}$C
            & $P$6$_{3}$22 (182)
            & 
                \begin{tabular}{c} 
                    $a$ = $b$ = 4.677 \\
                    $c$ = 4.344 \\
                \end{tabular}
              & 
                \begin{tabular}{c c c c} 
                    Fe = & 0.3196 & 0 & 0 \\
                    C = & 1/3 & 2/3 & 3/4 \\
                \end{tabular} \\
        \hline
        hexa-Co$_{3}$C
            & $P$6$_{3}$22 (182)
            & 
                \begin{tabular}{c} 
                    $a$ = $b$ = 4.5497 \\
                    $c$ = 4.4095 \\
                \end{tabular}
              & 
                \begin{tabular}{c c c c} 
                    Co = & 0.32348 & 0 & 0 \\
                    C = & 1/3 & 2/3 & 3/4 \\
                \end{tabular} \\
        \hline
        ortho-Fe$_{3}$C
            & $Pnma$ (62)
            & 
                \begin{tabular}{c} 
                    $a$ = 5.0510 \\
                    $b$ = 6.6867 \\
                    $c$ = 4.4906 \\
                \end{tabular}
            & 
                \begin{tabular}{c c c c} 
                    Fe$^{g}$ = & 0.1765 & 0.0664 & 0.3322 \\
                    Fe$^{s}$ = & 0.0337 & 0.25 & 0.8392 \\
                    C = & 0.8762 & 0.25 & 0.4381 \\
                \end{tabular} \\
        \hline
        ortho-Co$_{3}$C
            & $Pnma$ (62)
            & 
                \begin{tabular}{c} 
                    $a$ = 4.996 \\
                    $b$ = 6.614 \\
                    $c$ = 4.442 \\
                \end{tabular}
            & 
                \begin{tabular}{c c c c} 
                    Co$^{g}$ = & 0.1805 & 0.0683 & 0.3314 \\
                    Co$^{s}$ = & 0.0372 & 0.25 & 0.8451 \\
                    C = & 0.8794 & 0.25 & 0.4479 \\
                \end{tabular} \\
        \hline
        \hline
    \end{tabular}
\end{table*}

\begin{figure}[t]
    \centering
    \includegraphics[trim = 40 0 20 0, clip,width=1.0\columnwidth]{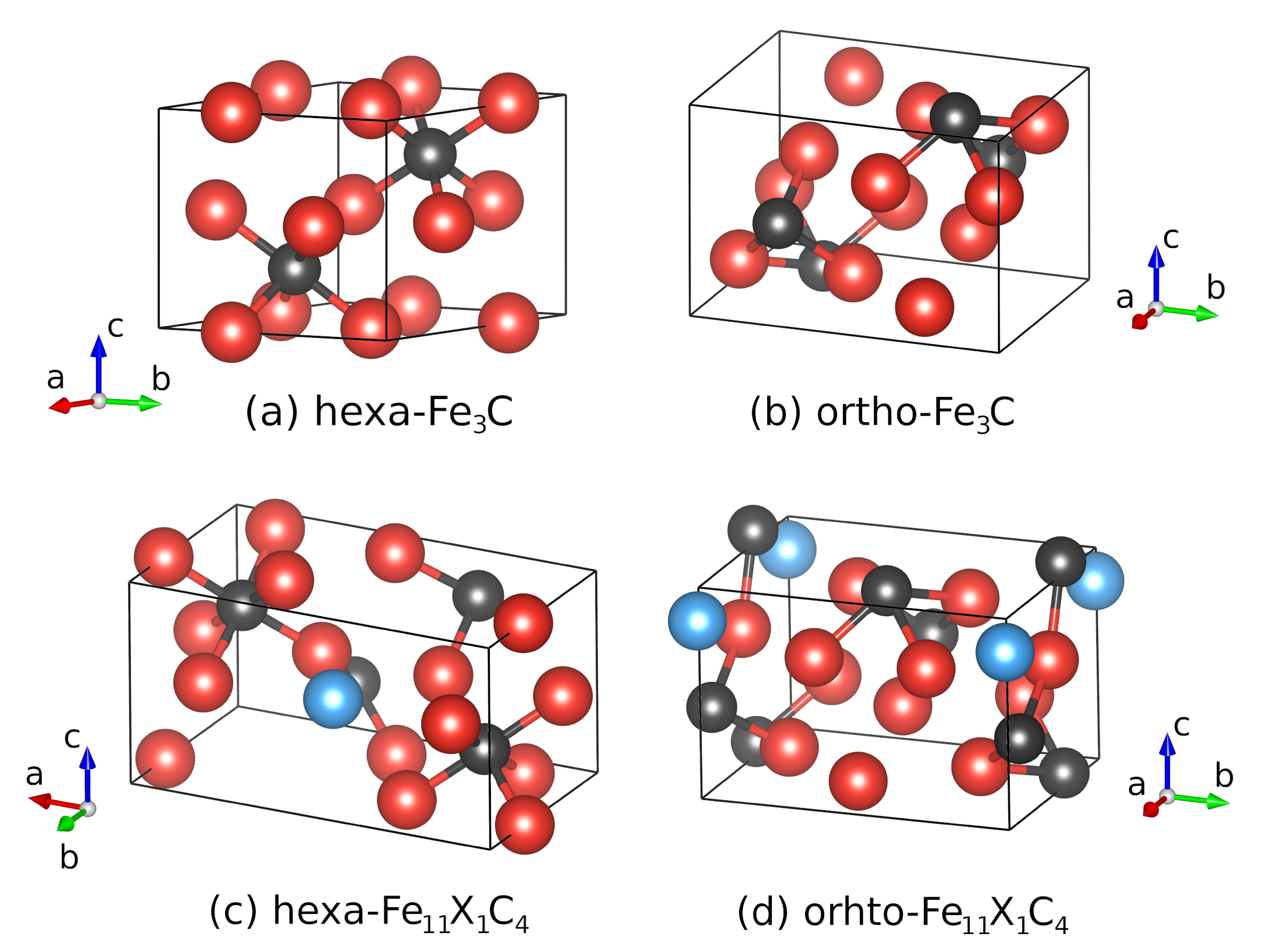}   
    \caption{
    Crystal structures of 
(a)~hexa-Fe$_{3}$C and 
(b)~ortho-Fe$_{3}$C, 
    together with corresponding structures of~(c,d) hexa- and ortho-Fe$_{11}$X$_{1}$C$_{4}$ alloys substituted with transition metal element X.
    Iron, carbon, and transition metal atoms are represented by red, black, and blue spheres, respectively.
}
\label{fig:pure_structures}
\end{figure}

%-------FPLO, GGA, k-mesh, delta d------ 
%
The calculations were done using density functional theory (DFT) implemented in the full-potential local-orbital code (FPLO18.00-52)~\cite{koepernikFullpotentialNonorthogonalLocalorbital1999}.
The FPLO includes relativistic effects in the full four-component formalism~\cite{opahle1999full}. 
Unless otherwise specified, the generalized gradient approximation (GGA) proposed by Perdew, Burke, and Ernzerhof (PBE)~\cite{perdewGeneralizedGradientApproximation1996} was used. 
For Brillouin zone integration, a tetrahedron method was used, with a mesh of $k$-points set to 40~$\times$~40~$\times$~40. 
All calculations were fully converged with a density criterion of 10$^{-7}$.

%--------structural parameters for Fe3C i Co3C, geometry optimization----
%
The initial structural data for hexagonal $\epsilon$-Fe$_{3}$C  
and orthorhombic $\theta$-Fe$_{3}$C phase were taken from experiment~\cite{yakel1985crystal,fruchart1984etudes}, see Table~\ref{tab:struct} and Fig.~\ref{fig:pure_structures}.
Hexa- and ortho-Co$_{3}$C structures were prepared using initial data for Fe$_3$C phases.
The volumes of all considered structures were optimized.
For hexagonal structures, the ratio of $c/a$ lattice parameters was optimized as well.
Wyckoff's positions were optimized in each system using forces with convergence criterion set to 10$^{-3}$~eV\,\AA{}$^{-1}$ within a scalar-relativistic approach and with spin polarization, see Table~\ref{tab:struct}. 
The crystal structures were visualized with the VESTA program~\cite{mommaVESTAThreedimensionalVisualization2008}.

%-------MAE, Vxc, FSM------------
%
To determine the magnetocrystalline anisotropy energy (MAE), following the completion of self-consistent scalar-relativistic calculations, single iterations of fully relativistic calculations were conducted for orthogonal directions of the magnetization.
To analyze the effect of the form of the exchange-correlation potential on the magnetic moment and MAE, the von~Barth and Hedin (vBH) \cite{vBH1972local}, Perdew and Zunger (PZ) \cite{PerdewZunger1981self}, Perdew and Wang (PW92) \cite{PerdewWang1992accurate}, and exchange-only (x-only) forms of the local spin density approximation (LSDA) were used in addition to the Perdew-Burke-Ernzerhof (PBE) \cite{perdewGeneralizedGradientApproximation1996} potential used throughout the paper.
In order to ascertain the dependency of the MAE on the magnetic moment, a fully relativistic approach was employed, utilizing of the fixed-spin-moment (FSM) method.

%-------VCA-------------
%
The virtual crystal approximation (VCA) was used to model the (Fe,Co)$_{3}$C alloys and alloying with B and N at the C position.
For the pseudobinary (Fe,Co)$_{3}$(B,C,N) alloys, a total of 231 compositions for both hexagonal and orthorhombic phase was prepared and evaluated.
In order to calculate structures with different percentages of Fe and Co, the lattice parameters and Wyckoff positions were averaged proportionally according to the Fe and Co content of the structure.

%------supercels-----------
%
To consider $\frac{1}{12}$ substitution of the Fe atoms with 3$d$, 4$d$, and 5$d$ elements, the symmetry of both unit cells was reduced, see Fig.~\ref{fig:pure_structures}. 
In the case of hexa-Fe$_{11}$X$_{1}$C$_{4}$ alloys, the space group was reduced to $P$121,
while for ortho-(Fe/Co)$_{11}$X$_{1}$C$_{4}$ alloys, the space group was reduced to $P$1$m$1. 
For supercells, the mesh of $k$-points was set to 25~$\times$~25~$\times$~25, while the charge density criterion was 10$^{-7}$. 
In all supercells, the lattice parameters were taken as they were in the initial optimized compounds, whereas the Wyckoff positions were optimized using forces.
The choice of a substitution concentration of $\frac{1}{12}$ is based on the trade-off between the relatively low concentration value and the computation time. 
Within the supercell method, an alternative concentration of $\frac{1}{24}$ would have to be based on twice the elemental cell, which would increase calculation time by several times.

%---------formation energy----------
%
Formation energies ($E_f$) of Fe$_{11}$X$_{1}$C$_{4}$ alloys were calculated from equation:
\begin{equation}
E_f = E_{Fe_{11}X_1C_{4}} - 11 E_{Fe} - E_X - 4 E_C,
\label{eq:formation_energy}
\end{equation}
where $E_{Fe_{11}X_1C_{4}}$, $E_{Fe}$, $E_C$, and $E_X$ are the total energies of the $E_{Fe_{11}X_1C_{4}}$ supercell and the crystals of iron, carbon (graphite), and transition metal (X), respectively. 
To determine the total energies, we prepared and optimized the geometry of 3$d$, 4$d$, and 5$d$ crystals.
The formation energies of hexa-Co$_{11}$X$_{1}$C$_{4}$ compositions were calculated analogically.
The formation energies are determined for the ground state, that is, at 0~K.

%--------MAE calculations-------
%
%---MAE for hexa---
%
For the hexagonal crystal system, the MAE can be estimated from the difference in energies calculated for magnetization along the unique crystal axis and the axis perpendicular to it.
It can give both non-negative and negative values, for perpendicular and in-plane magnetocrystalline anisotropy, respectively.

%-----MAE for ortho---
%
Following our approach for calculating the MAE in orthorhombic structures~\cite{snarski2024searching}, we have applied the same method in this work.
The axis of easy magnetization is defined by the lowest energy of the three fully relativistic quantization directions along the main crystal axes ([100], [010], and [001] along $a$, $b$, and $c$, respectively). 
For each composition, we define an axis with the lowest energy (E$_1$), medium energy (E$_2$), and highest energy (E$_3$).
The MAE of an orthorhombic system is determined by the difference between the medium and lowest energy, which gives us only non-negative MAE values.
To complete the determination of MAE in the orthorhombic structure, we introduced the DE$_{32}$ parameter describing the anisotropy energy between E$_3$ and E$_2$. 
For the experimental determination of the effective MAE, similar approach was used by Zhdanova~\textit{et al.}~\cite{zhdanova_magnetic_2013}.

%-------magnetic hardness---------
%
Another key parameter when discussing permanent magnet  applications is magnetic hardness~\cite{coey_perspective_2020}, which can be expressed as :
\begin{equation}
    \kappa=\sqrt{\frac{K}{\mu_{0}M_{\rm{s}}^{2}}},
\end{equation} 
where $K$ is magnetocrystalline anisotropy constant (interpreted as MAE), $\mu_0$ is vacuum permeability, and $M_{\rm{s}}$ is saturation magnetization.
The latter might be given as: 
\begin{equation}
    M_{\rm{s}}=\frac{m_{\rm{s}}+m_{\rm{l}}}{V},
\end{equation}
where $m_{\rm{s}}$ is total spin magnetic moment, $m_{\rm{l}}$ is total orbital magnetic moment, and $V$ is unit cell volume. 
While $\kappa > 1$ denotes hard magnetic materials, it is assumed that $0.1 < \kappa < 1$ defines the range of semi-hard magnetic materials~\cite{coey_perspective_2020}.
To use the semi-hard materials as permanent magnets, the condition that  $\kappa \gtrsim 0.5$ should be met~\cite{coey_perspective_2020}.

%---------Curie temperature from DLM
%
Curie temperature ($T_\mathrm{C}^{\mathrm{MFT}}$) of (Fe$_{0.916}$X$_{0.084}$)$_3$C alloys with the 3$d$, 4$d$, and 5$d$ transition metals (X) were calculated using mean-field theory~\cite{gyorffy_first-principles_1985, kudrnovsky_exchange_2004}.
Here, mean-field theory is an approximation for fluctuations in configuration of magnetic moment orientations, where all interactions with any body are replaced by an averaged interaction~\cite{gyorffy_first-principles_1985}.
When calculating the mean-field Curie temperature ($T_\mathrm{C}^{\mathrm{MFT}}$), a coherent potential approximation (CPA) procedure
 can be used to average and iterate to self-consistency~\cite{soven_coherent-potential_1967,gyorffy_first-principles_1985}.
The (Fe$_{0.916}$X$_{0.084}$)$_3$C concentration, modeled here with CPA, was chosen to match the Fe$_{11}$X$_{1}$C$_{4}$ concentration, modeled with supercell method.
We used the:
\begin{equation}\label{eq:Curie_T}
k_\mathrm{B} T_\mathrm{C}^{\mathrm{MFT}} = \frac{2}{3} \frac{E_{\mathrm{DLM}} - E_{\mathrm{FM}}}{c},
\end{equation}
equation.
Here, $E_{\mathrm{DLM}}$ and $E_{\mathrm{FM}}$ 
are the total energies of the paramagnetic and ferromagnetic solutions, $k_{\mathrm{B}}$ is Boltzmann's constant, and $c$ is the number of \textit{magnetic} atoms in formula. 
For Fe$_3$C we assume $c$ equal 3.
To model the paramagnetic configuration, 
we used the disordered local moment (DLM) approach 
~\cite{heine_theory_1981} 
with the coherent potential approximation (CPA) 
~\cite{soven_coherent-potential_1967}.
In the DLM approach, the local magnetic moments can rotate, and the self-consistent calculations are conducted for the given arrangement of them.
Numerous magnetic moment orientations are calculated and averaged.
The paramagnetic state is determined by taking the total magnetic moment equal to zero.
A ferromagnetic (FM) ground state, on the other hand, is defined as an equilibrium configuration of parallel collinear magnetic moments.
Scalar-relativistic DLM-CPA calculations were conducted using the FPLO5 code, which is the most recent version of the FPLO code that incorporates CPA. 
However, the FPLO5 code lacks the PBE-GGA implementation, forcing the use of the LDA-PW92 exchange-correlation potential for CPA calculations~\cite{PerdewWang1992accurate}.

\begin{figure}[t]
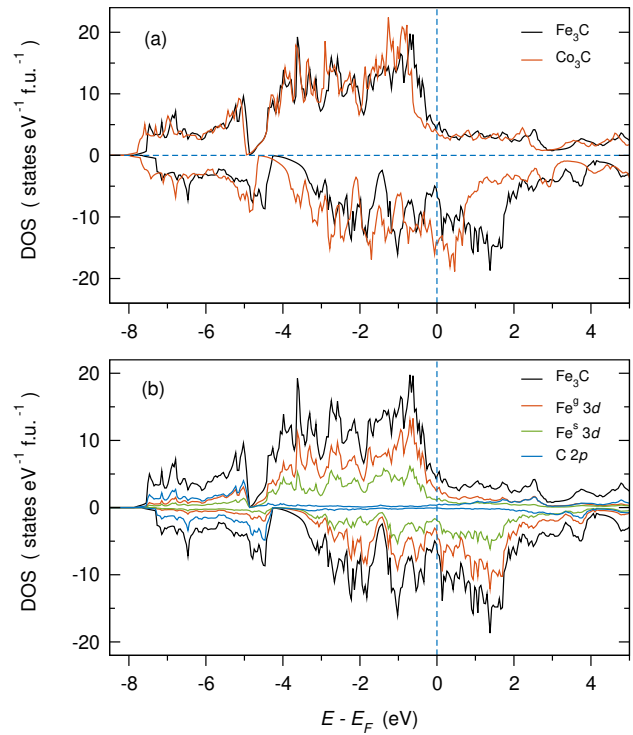

    \centering
    \includegraphics[trim = 0 40 0 0, clip,width=0.95\columnwidth]{fe3c_co3c_DOS.eps}
    \includegraphics[clip,width=0.95\columnwidth]{fe3c_ortho_DOS.eps}
    \caption{
Densities of states (DOS) for ortho-Fe$_{3}$C and ortho-Co$_{3}$C.
The calculations were performed with the FPLO18 code using the PBE exchange-correlation potential.
}
\label{fig:DOS}
\end{figure}

\begin{figure*}[t]
    \centering
    \includegraphics[trim = 6 6 7 6, clip,width=1\textwidth]{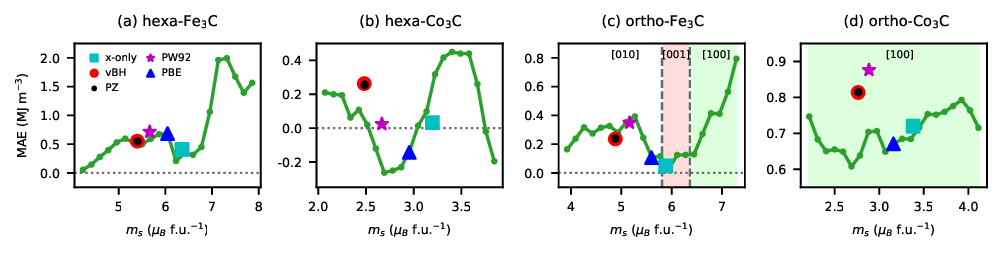}   
    \caption{
    The dependence of the magnetocrystalline anisotropy energy (MAE) on the spin magnetic moment ($m_{s}$) for 
    (a)~hexa-Fe$_{3}$C, 
    (b)~hexa-Co$_{3}$C, 
    (c)~ortho-Fe$_{3}$C, and
    (d)~ortho-Co$_{3}$C. 
    The calculations were performed with the FPLO18 code using the Perdew-Burke-Ernzerhof (PBE) exchange-correlation potential.
    Equilibrium values of MAE and $m_{s}$ were obtained using functionals of von~Barth and Hedin (vBH), Perdew and Zunger (PZ), Perdew and Wang 92 (PW92), and LDA exchange-only (x-only), besides Perdew-Burke-Ernzerhof (PBE) in the generalized gradient approximation (GGA). 
    They are represented by symbols.
    The structures with different axes of easy magnetization are distinguished by different background colors: white for [010], light red for [001], and light green for [100].
    In the case of  ortho-Co$_{3}$C there is one easy axis for the whole considered range of magnetic moments, namely [100].
}
\label{fig:FSM_Vxc}
\end{figure*}

\begin{figure}[t]
    \centering \includegraphics[trim = 7 5 5 5,clip,width=\columnwidth]{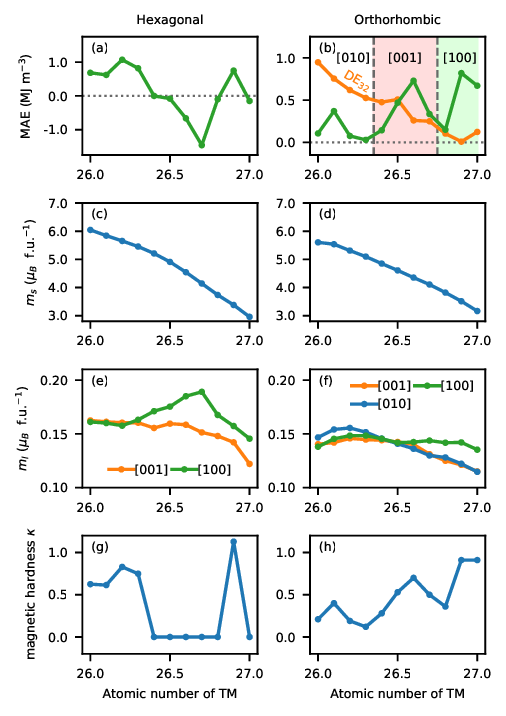}
    \caption{
    Intrinsic properties calculated for hexa- and ortho-(Fe$_{1-x}$Co$_{x}$)$_{3}$C alloys.
    The increase in Co concentration $x$ is expressed as an increase in the atomic number of the transition metal (TM) element; from 26 (Fe) to 27 (Co).
    (a, b)~The magnetocrystalline anisotropy energy (MAE) is represented by the green line with markers, and the energy difference between the two higher energies of three energies calculated for three orthogonal magnetization directions (DE$_{32}=E_{3}-E_{2}$) by the orange line with markers.
    (c-f)~The total spin ($m_{s}$) and orbital ($m_{l}$) magnetic moments, and (g, h)~the magnetic hardness ($\kappa$). 
    The concentration range for panel (b) is divided into regions characterized by the magnetic easy axes.
    Calculations were performed with the virtual crystal approximation (VCA) using the FPLO18 code with PBE exchange-correlation potential.
    }
    \label{fig:VCA_1D_MAE_ms_ml}
\end{figure}

\begin{figure}[t]

\centering 
\includegraphics[trim = 0 0 0 0,clip,height=0.385\columnwidth]{hexa_feco3c_tc_vs_x.eps}
\includegraphics[trim = 50 0 0 0,clip,height=0.385\columnwidth]{ortho_feco3c_tc_vs_x.eps}
\caption{
The mean-field-theory Curie temperatures ($T_{\mathrm{C}}^{\mathrm{MFT}}$) calculated for (a) hexa- and (b) ortho-(Fe$_{1-x}$Co$_{x}$)$_{3}$C alloys as a function of Co concentration.
The results were obtained using the FPLO5 code and the Perdew and Wang (PW92) functional.
The chemical disorder was modeled with coherent potential approximation (CPA). 
The paramagnetic state was modeled using the disorder local moment (DLM-CPA) method.
}
\label{fig:tc_ortho_feco3c}
\end{figure}

\begin{table}[t]
\center
\caption{
Calculated magnetic properties of hexagonal and orthorhombic (Fe$_{1-x}$Co$_{x}$)$_{3}$C alloys.
Spin and orbital magnetic moments [$m_{s}$ and $m_l$ ($\mu_\mathrm{B}$ per atom or f.u.)],
magnetocrystalline anisotropy energy [MAE~(MJ\,m$^{-3}$)], and
magnetic hardness [$\kappa$].
Magnetic moments on transition-metal (TM) elements for ortho-(Fe$_{1-x}$Co$_{x}$)$_{3}$C alloys are given for both general and special non-equivalent crystallographic sites.
The increase in Co concentration $x$ is expressed as an increase in the atomic number of the transition metal (TM) element; from 26.0 for Fe to 27.0 for Co.
The calculations were performed with the FPLO18 code using the PBE potential and virtual crystal approximation.
}
\label{tab:1D_VCA}
\begin{tabular}{c c c c c c}
        \hline
        \hline
        \multirow{2}{*}{Hexagonal} & \multicolumn{5}{c}{Atomic number of TM} \\
        \cline{2-6}
        & 26.0 & 26.2 & 26.7 & 26.9 & 27.0 \\
        \hline
        %\hline
        $m_{s}$(Fe/Co) & 2.12 & 1.97  & 1.44  & 1.17  & 1.03 \\
        %\hline
        $m_{s}$(C) & -0.30 & -0.27 & -0.18  & -0.14  &  -0.12 \\
        %\hline
        total $m_{s}$ & 6.04 & 5.65 & 4.14  & 3.38  &  2.95 \\
        %\hline
        $m_{l}$(Fe/Co) & 0.053 & 0.052 & 0.049  & 0.046  & 0.040 \\
        %\hline
        $m_{l}$(C) & 0.00 & 0.00 & 0.00  & 0.00  & 0.00 \\
        %\hline
        total $m_{l}$ & 0.16 & 0.16 & 0.15  & 0.14  & 0.12 \\
        %\hline
        MAE & 0.68 & 1.07 & -1.46 & 0.75 & -0.15 \\
        $\kappa$ & 0.63 & 0.83 & 0.00 & 1.13 & 0.00 \\
        \hline
        \multirow{2}{*}{Orthorhombic} & \multicolumn{5}{c}{Atomic number of TM} \\
        \cline{2-6}
        & 26.0 & 26.1 & 26.6 & 26.9 & 27.0 \\
         %& \multicolumn{5}{ |c| }{Atomic number of TM} \\
        %\hline
         %& 26.0 & 26.1 & 26.6 & 26.9 & 27.0 \\
        \hline
        %\hline
        $m_{s}$(Fe$^{g}$/Co$^{g}$) & 1.93 & 1.90  & 1.52  & 1.24  & 1.12 \\
        %\hline
        $m_{s}$(Fe$^{s}$/Co$^{s}$) & 2.01 & 1.99 & 1.47  & 1.14  & 1.02 \\
        %\hline
        $m_{s}$(C) & -0.28 & -0.26 & -0.18  & -0.12  & -0.11 \\
        %\hline
        total $m_{s}$ & 5.56 & 5.53 & 4.34  & 3.51  & 3.16 \\
        %\hline
        $m_{l}$(Fe$^{g}$/Co$^{g}$) & 0.042 & 0.044 &  0.038 & 0.035  & 0.034 \\
        %\hline
        $m_{l}$(Fe$^{s}$/Co$^{s}$) & 0.054 & 0.055 & 0.050  & 0.048  & 0.043 \\
        %\hline
        $m_{l}$(C) & 0.00 & 0.00 & 0.00  & 0.00  & 0.00 \\
        %\hline
        total $m_{l}$ & 0.14 & 0.14 & 0.14  & 0.12  & 0.12 \\
        %\hline
        MAE & 0.11 & 0.37 & 0.73 & 0.82 & 0.67 \\
        $\kappa$ & 0.21 & 0.40 & 0.70 & 0.91 & 0.91 \\
        \hline
        \hline
    \end{tabular}
\end{table}

\begin{table}[t]
\center
\caption{
Total spin magnetic moments [$m_{s}$ ($\mu_\mathrm{B}$\,f.u.$^{-1}$)] and
the magnetocrystalline anisotropy energies [MAE (MJ\,m$^{-3}$)] calculated
for hexa-Fe$_{3}$C, hexa-Co$_{3}$C, ortho-Fe$_{3}$C, and ortho-Co$_{3}$C. 
The equilibrium values were obtained using the von~Barth and Hedin (vBH), Perdew and Zunger (PZ), Perdew and Wang 92 (PW92), Perdew-Burke-Ernzerhof (PBE), and LDA exchange-only (x-only) exchange-correlation potentials (V$_{xc}$). 
The calculations were performed with the FPLO code.
At the first line, the VASP-PAW-PBE results 
from Ref.~\cite{wu2017exploring} are presented. 
}
\label{tab:Vxc}
\begin{tabular}{ccccccccc}
        \hline
        \hline
         \multirow{2}{*}{V$_{xc}$} &  \multicolumn{2}{c}{hexa-Fe$_{3}$C} & \multicolumn{2}{c}{hexa-Co$_{3}$C} & \multicolumn{2}{c}{ortho-Fe$_{3}$C} & \multicolumn{2}{c}{ortho-Co$_{3}$C} \\
         \cline{2-9}
            &  $m_{s}$ & MAE & $m_{s}$ & MAE & $m_{s}$ & MAE & $m_{s}$ & MAE\\
        \hline
        PBE \cite{wu2017exploring}
            & 5.92  & 0.57 & 2.76 & 0.02 & 5.52 & 0.05 & 3.12  & 0.81 \\
        PBE %($m_{s(GGA)}$, MAE$_{GGA}$) 
            & 6.05  & 0.68 & 2.95  & -0.15 & 5.56  & 0.11 & 3.16  & 0.67\\
        vBH %($m_{s(vBH)}$, MAE$_{vBH}$) 
            & 5.40  & 0.55 & 2.48  & 0.26 & 4.88  & 0.24 & 2.76  & 0.81\\
        PZ %($m_{s(PZ)}$, MAE$_{PZ}$) 
            & 5.40 & 0.55 & 2.48  & 0.26 & 4.88 & 0.24 & 2.77  & 0.81\\
        PW92 %($m_{s(PW92)}$, MAE$_{PW92}$) 
            & 5.66  & 0.71 & 2.67  & 0.02 & 5.16  & 0.35 & 2.89  & 0.88\\
        x-only %($m_{s(Exch)}$, MAE$_{Exch}$) 
            & 6.36  & 0.41 & 3.20  & 0.03 & 5.88  & 0.05 & 3.38  & 0.72\\
        \hline
        \hline
    \end{tabular}
\end{table}

\section{Results and Discussion}

%-------short introduction for the section------------
%
This paper consists of four parts. 
In the first, we discuss Fe$_3$C and Co$_3$C compounds. 
In the second, the (Fe,Co)$_3$C alloys. 
In the third, transition metal substitutions in Fe$_3$C and Co$_3$C compounds.
In the fourth, B and N substitutions at the C position in (Fe,Co)$_3$C alloys.
Each time, we consider both hexagonal and orthorhombic phases.

\subsection{Fe$_{3}$C and Co$_{3}$C compounds}

%---------short introduction to the subsection-----
%
In this section, dedicated to the Fe$_3$C and Co$_3$C compounds, we will present their densities of states, examine how the choice of the exchange-correlation potential affects the obtained values of magnetic moments and magnetocrystalline anisotropy energy (MAE), and study the dependence of MAE on the magnetic moment. 
We started this study from the geometry optimization of considered phases, and the optimized structural parameters are presented in Table~\ref{tab:struct}.

%--------formation energy and phase stability-----
%
The experimental standard enthalpy of formation ($\Delta H_f^0$) 
for Fe$_3$C is $4.7\pm1.1$~kJ\,mol$^{-1}$ ($0.049\pm0.011$~eV\,atom$^{-1}$)
and 
for Co$_3$C is $2.4\pm1.1$~kJ\,mol$^{-1}$ ($0.025\pm0.011$~eV\,atom$^{-1}$)~\cite{meschel_standard_1997}.
Positive enthalpy values indicate instability.
Similarly, the previously calculated corresponding formation energies are positive and are 0.016~eV\,atom$^{-1}$~\cite{ande2012first} and 0.07~eV\,atom$^{-1}$~\cite{konyaeva_electronic_2009} for ortho-Fe$_3$C and
0.05~eV\,atom$^{-1}$ for ortho-Co$_3$C~\cite{marshall_high-pressure_2021}.
The formation energies calculated in this work are positive for all four compounds considered in this section and equal to 0.085~eV\,atom$^{-1}$ for hexa-Fe$_3$C, 0.70~eV\,atom$^{-1}$ for hexa-Co$_3$C, 0.055~eV\,atom$^{-1}$ for ortho-Fe$_3$C, and 0.14~eV\,atom$^{-1}$ for ortho-Co$_3$C.
Orthorhombic phases have lower energies than hexagonal ones, suggesting that Fe$_3$C and Co$_3$C will prefer the cementite-type structure.
The formation energies are also relatively low, giving the possibility of stabilizing the compounds with additional factors such as alloying or the presence of a second phase.
Nanoparticles of ortho-Co$_3$C were successfully synthesized~\cite{carroll_magnetic_2012, martinez-teran_nucleation_2019} and the bulk form of ortho-Co$_3$C has been obtained under high pressure (above 4.8~GPa)~\cite{marshall_high-pressure_2021}.
Another method of stabilizing ortho-Co$_3$C is its alloying with a stable isostructural Co$_3$B phase~\cite{zieschang_magnetic_2019}.

%-----------Curie temperature-------------
%
%--------intro TC + expt values--------------
%
For the application of the material in permanent magnets, it is important that its Curie temperature far exceeds the room temperature.
%
%---------hexa-Fe3C+X---------
%
For metastable hexa-Fe$_3$C, 
the Curie temperature is difficult to determine 
and only an approximate experimental value of $T_C \geq 653$~K is given~\cite{barinov_carbonization_2014}.
Whereas, off-stoichiometric hexa-Fe$_{2.2}$C has $T_C$ of about 723~K~\cite{le_caer_characterization_1982}.
The calculated $T_C$ for hexa-Fe$_3$C is 1040~K, which is much higher than the mentioned experimental estimation ($T_C \geq 653$~K).
The observed difference in $T_C$ may come from both the known overestimation of $T_C$ in the mean-field theory and the underestimation of the experimental value.
%
%---------hexa-Fe3C+X---------
%
Because, as our calculations indicate, the hexa-Co$_3$C phase has a much higher formation energy than the other phases under consideration, it is much more difficult to stabilize, and we have not found literature experimental results on both its synthesis and Curie temperature.
The $T_C$ value we determined for hexa-Co$_3$C is 310~K.

%---------ortho-Fe3C+X---------
The experimental Curie temperatures of ortho-Fe$_3$C is equal to about 481~K~\cite{nicholson_solubility_1957} or 483~K~\cite{tsuzuki_high_1984}, while our calculations indicate 570~K,
overestimating the experimental value by about 20\%, which is the typical behavior of the mean-field $T_C$ results~\cite{turek_ab_2003,ebert_calculating_2011,ke_effects_2013}.
%
%---------ortho-Co3C+X---------
%
The experimental Curie temperature for ortho-Co$_3$C can be estimated from the results for ortho-Co$_3$C-based nanoparticles, but these are ambiguous and depend on the specific properties of the nanoparticles.
For ortho-Co$_3$C-based nanoparticles, we encounter the following Curie temperatures: 
498~K~\cite{mikhalev_magnetic_2019}, 
510~K~\cite{harris_high_2010}, 
563~K~\cite{turgut_metastable_2016}, 
and 650~K~\cite{el-gendy_enhanced_2014}.
$T_C$ calculated by us for ortho-Co$_3$C is equal to 458~K.
The observed discrepancy between the mean-field result and measurements confirms the significant effect of nanostructuring on the magnetic properties of ortho-Co$_3$C nanoparticles.

%---------DOS of ortho-Fe3C and ortho-Co3C-----
%
Figure~\ref{fig:DOS} presents the densities of states (DOS) calculated for orthorhombic Fe$_3$C and Co$_3$C.
The valence band of ortho-Fe$_3$C is dominated by Fe~3$d$ and C~2$p$ orbitals, between which a clear hybridization is visible.
The spin polarization of the valence band is noticeably stronger for Fe$_3$C.
For densities of states for hexagonal phases, see \textit{Supplementary Material}.

%----------spin magnetic moments-------------
%
The total spin magnetic moments, corresponding to the spin polarization in the DOS, are 5.56 and 3.16~$\mu_\mathrm{B}$\,f.u.$^{-1}$ for ortho-Fe$_3$C and ortho-Co$_3$C, respectively.
The values of calculated total and partial spin magnetic moments are summarized in the Table~\ref{tab:1D_VCA}.
In the Table~\ref{tab:literature}, we have further compared our results (obtained with the full-potential PBE approach) with those from the literature.
Given the differences in the calculation methods used, we rate the level of agreement as good.
Later, we will discuss how the choice of exchange-correlation potential affects the spin magnetic moment.

%---ms-----porownanie wynikow ms uzyskanych dla Fe/Co3C z danymi literaturowymi dla bulkowych bcc-Fe oraz fcc-Co-----------------------
%
The spin magnetic moment on the Fe atom, determined in our PBE calculations, shows a higher value for hexa-Fe$_{3}$C (2.12~$\mu_\mathrm{B}$) and lower values for ortho-Fe$_{3}$C phase (1.93 and 2.01~$\mu_\mathrm{B}$ for general and special Fe positions, respectively).
All values are smaller than the magnetic moments for bcc Fe (2.17~$\mu_\mathrm{B}$~\cite{wu1999spin} or 2.20~$\mu_\mathrm{B}$~\cite{jiang2003carbon}).
In contrast, the hexa-Co$_{3}$C phase exhibits smaller spin magnetic moments on Co atoms (1.02~$\mu_\mathrm{B}$) than the ortho-Co$_{3}$C phase (1.02 and 1.12~$\mu_\mathrm{B}$ for special and general Co positions, respectively).
The $m_{s}$ values for bulk fcc Co are higher (1.66~$\mu_\mathrm{B}$~\cite{wu1999spin}; 1.58~$\mu_\mathrm{B}$~\cite{budko2019calculation}) than those calculated for Co$_{3}$C.

%----------MAE values for Fe3C and Co3C compounds
%
In the present work, our aim is not to repeat the calculation of magnetic moments, which, as we see in the Table~\ref{tab:literature}, has already been done, but to calculate the  magnetocrystalline anisotropy energies (MAE) and determine magnetic hardness of considered materials.
The MAE values calculated with PBE functional (at 0~K) are 0.68, -0.14, 0.11, and 0.67~MJ\,m$^{-3}$ for hexa-Fe$_3$C, hexa-Co$_3$C, ortho-Fe$_3$C and ortho-Co$_3$C, respectively, see Table~\ref{tab:Vxc}.
%
%Of these values, the result for ortho-Co$_3$C is noteworthy, since indicates a magnetic hardness of 1.03, thereby establishing its  membership in the class of magnetically hard materials ($\kappa > 1$).
%
The calculated MAE of ortho-Co$_3$C is in agreement with experimental magnetocrystalline anisotropy constant determined for ortho-Co$_3$C nanoparticles as 0.74$\pm$0.1~MJ\,m$^{-3}$~\cite{el-gendy_enhanced_2014}.
The corresponding MAE values calculated within projector augmented wave PBE method are 
0.57, 0.02, 0.05, and 0.81~MJ\,m$^{-3}$ for hexa-Fe$_3$C, hexa-Co$_3$C, ortho-Fe$_3$C and ortho-Co$_3$C, respectively~\cite{wu2017exploring}, please check Table~\ref{tab:Vxc}.
The similarity between two sets of computational results is qualitatively good, whereas the quantitative differences come most probably from the shape approximation of the potential used by Wu~\textit{et al.}~\cite{wu2017exploring}.
Another theoretical approach determines MAE of ortho-Co$_3$C as between 0.8 and 0.9~MJ\,m$^{-3}$~\cite{el-gendy_enhanced_2014}, identifying ortho-Co$_3$C phase as hard magnetic.
Since all the mentioned results are calculated for a temperature of 0~K and additionally depend on the approximation used, they should be interpreted with due caution.
In order to establish a context for the PBE-MAE results, we will present an analysis of the impact of the form of exchange-correlation potential on the MAE values and the relation between values of MAE and spin magnetic moment.

%-------------various Vxc------------
%
In our previous works, we demonstrated that the calculated MAE values for CeFe$_{12}$ and MnB depend on the choice of the exchange-correlation potential~\cite{snarski2022effect, snarski2024searching}. 
We also showed there, that the MAE is correlated with the spin magnetic moment ($m_s$).
Here, we performed MAE and $m_s$ calculations for the Fe$_{3}$C and Co$_{3}$C compounds using five forms of exchange-correlation potential: Perdew-Burke-Ernzerhof (PBE), von~Barth and Hedin (vBH), Perdew and Zunger (PZ), Perdew and Wang 92 (PW92), and LSDA exchange only (x-only).
The self-consistent equilibrium results of MAE and $m_s$ are shown in Fig.~\ref{fig:FSM_Vxc}, and in Table~\ref{tab:Vxc}.
For each phase, the calculated magnetic moment increases going from vBH and PZ, through PW92, PBE, to the x-only potential.
In the case of MAE, there is a scattering of results with a range of up to about 0.4~MJ\,m$^{-3}$.
However, we do not observe any clear trends in the MAE values.
In contrast, the magnetic moments increase in each case with the following order of exchange-correlation potentials: vBH/PZ~$\rightarrow$ PW92~$\rightarrow$ PBE~$\rightarrow$ \textit{x-only}.
In the case of referred ortho-Co$_3$C, alternative functionals result in MAE values that are even higher than those obtained with PBE. 
Thus, ortho-Co$_3$C appears to be a promising hard-magnetic material, regardless of the approximation used.

%--------fixed spin moment--------
%
Furthermore, we performed calculations of the dependencies of MAE on the fixed spin magnetic moment (FSM), see Fig.~\ref{fig:FSM_Vxc}.
We observe, that the dependencies of MAE on fixed spin magnetic moment overlap with equilibrium data points calculated with different functionals.

%----- hexa-Fe$_{3}$C -----
%
For hexa-Fe$_{3}$C, in the whole range of magnetic moments, we observe positive MAE values, implying uniaxial magnetocrystalline anisotropy.
The MAE ranges from about 0 to 2~MJ\,m$^{-3}$.
Unfortunately, while lowering the magnetic moment of the system is a relatively straightforward task, engineering an increase in the magnetic moment of Fe$_3$C can be difficult.
It is because Fe$_3$C has one of the highest magnetic moments among isostructural compounds, and Fe substitution by both Mn and Co leads to a lower magnetic moment.

%---- hexa-Co$_{3}$C -----
%
The magnetic moment of Co$_{3}$C phases is about half that of Fe$_{3}$C phases.
For hexa-Co$_{3}$C, with increasing magnetic moment, we observe both positive and negative values of MAE, which indicate changes between easy-axis and easy-plane types of magnetocrystalline anisotropy, respectively.
All MAE values determined for hexa-Co$_{3}$C phase classify it as a magnetically soft material.

%------ ortho-Fe$_{3}$C -------------
%
In the case of ortho-Fe$_{3}$C we observe somewhat lower equilibrium magnetic moments than for hexa-Fe$_{3}$C.
For ortho-Fe$_{3}$C, with increasing magnetic moment, the axis of easy magnetization undergoes a change from [010] to [001] and subsequently to [100].
The calculated equilibrium MAE values range from 0.05 to 0.35~MJ\,m$^{-3}$.
For the highest considered magnetic moments, above 7~$\mu_\mathrm{B}$\,f.u.$^{-1}$, the MAE increases up to 0.8~MJ\,m$^{-3}$.
Unfortunately, as we have already said for hexa-Fe$_3$C counterpart, raising the magnetic moment of ortho-Fe$_3$C might be a difficult task.
While previous experiments on Fe$_3$(B,C) borocarbides have demonstrated that substituting C by B can result in an approximate 10\% increase in the magnetic moment of the system~\cite{nicholson_solubility_1957}, this would still not be sufficient to reach the high MAE range.

%----- ortho-Co$_{3}$C-----
%
In the ortho-Co$_{3}$C system, the obtained PBE-MAE values vary only slightly with the change of the spin magnetic moment, from 0.6 to 0.8~MJ\,m$^{-3}$.
The equilibrium MAE values calculated in LDA (vBH, PZ, and PW92) exceed this range and reach 0.9~MJ\,m$^{-3}$.
As we said already in the previous section, the calculated MAE of ortho-Co$_{3}$C is in agreement with experimental magnetocrystalline anisotropy constant determined for ortho-Co$_3$C nanoparticles as 0.74$\pm$0.1~MJ\,m$^{-3}$~\cite{el-gendy_enhanced_2014}.
Since the observed MAE values are relatively high and the magnetic moments are low, the magnetic hardness is close or above one for the whole considered range of fixed spin magnetic moments.

%----section summary--------
%
In summary, from the perspective of magnetic hardness, the ortho-Co$_3$C phase deserves the most attention. 
Calculations suggest that magnetic hardness close to unity should persist not only at 0~K, but also at room temperature when the magnetic moment decreases.
In addition, previous experiments show high values of the magnetic anisotropy constant for ortho-Co3C nanoparticles and Curie temperatures above 500~K.
However, the relatively high price of cobalt works against the potential applications of this system.
Having determined the magnetic hardness of Fe$_3$C and Co$_3$C compounds, in the next section we will address the full ranges of intermediate concentrations of (Fe,Co)$_3$C alloys.

\subsection{(Fe,Co)$_{3}$C alloys \label{sec:feco3c}}

Since Fe and Co have consecutive atomic numbers of 26 and 27, their alloys can be modeled using a virtual atom with a fractional atomic number determined by the concentration of the elements.
Our goal is to study how the MAE of the system changes with Co concentration and identify promising compositions with elevated MAE.
In our attempt we are motivated by previous studies, such as for FeCo, (Fe,Co)$_2$B and (Fe,Co)$_{16}$C alloys~\cite{burkert2004giant, edstrom_magnetic_2015,marciniak_structural_2023}, which showed higher MAE values for intermediate compositions than for compounds at the extremes of the Fe/Co range.
Moreover, we will determine how the Curie temperature and the spin and orbital magnetic moments change with Co concentration.

%-------MAE hexa-(Fe,Co)3C-----
%
At first, we analyze the concentration dependence of MAE of the hexa-(Fe,Co)$_{3}$C phase, see Fig.~\ref{fig:VCA_1D_MAE_ms_ml}(a) and Table~\ref{tab:1D_VCA}.
We observe both positive and negative MAE values, indicating uniaxial and in-plane magnetocrystalline anisotropies, respectively. 
The highest MAE of 1.07~MJ\,m$^{-3}$ was obtained for a concentration of 20\%~Co, and the minimum MAE of -1.46~MJ\,m$^{-3}$ for 70\%~Co.
This trend is not surprising, as an analogous course of MAE($x$) with a positive maximum and a negative minimum was observed for the mentioned before (Fe$_{1-x}$Co$_{x}$)$_{2}$B and (Fe$_{1-x}$Co$_{x}$)$_{16}$C alloys~\cite{edstrom_magnetic_2015,marciniak_structural_2023}.
As argued in the indicated works on uniaxial alloys, the course of the MAE($x$) relationship is due to the filling of the valence band of the alloy as the Co concentration increases.
Since MAE is a subtle quantity that depends on the details of the band structure in the close vicinity of the Fermi level, an increase in valence band filling by even about 0.1 electron per atom can lead to significant changes in MAE.

%-------MAE ortho-(Fe,Co)3C --------
%
For ortho-(Fe,Co)$_{3}$C alloys, the Co concentration range is divided into regions: from 0\% to 35\%, from 35\% to 75\%, and from 75\% to 100\%, with the directions of magnetic easy axes [010], [001], and [100], respectively, see Fig.~\ref{fig:VCA_1D_MAE_ms_ml}(b).
The highest MAE value of 0.82~MJ\,m$^{-3}$ was obtained for a concentration of 90\%~Co.
The course of MAE($x$) clearly differs from the result for the hexagonal phase (Fe,Co)$_3$C and the mentioned tetragonal phases Fe-Co-B.
This is due, among other things, to the presence of three main crystallographic axes, where each of them can be an easy axis with a positive MAE value.
On the other hand, the details of the MAE($x$) waveform itself are due, as we mentioned above, to the gradual filling of the valence band as the Co concentration increases.
Additionally, for orthorhombic alloys, we show the energy difference between two higher energies (DE$_{32}=E_{3}-E_{2}$) complementing the MAE.
The low DE$_{32}$ values, as observed for Co-rich phases, strengthen the uniaxial character of the magnetocrystalline anisotropy of these systems.

Unfortunately, there is no systematic study of the course of the magnetocrystalline anisotropy constant on Co concentration in (Fe,Co)$_3$C alloys to verify our theoretical predictions.
From scattered information, we learn that Choe~\textit{et al.} obtained for the powder ortho-Fe$_3$C sample an MAE equal to 0.405~MJ\,m$^{-3}$ at 5~K~\cite{choe_easy_2016}.
Wu~\textit{et al.} synthesized a melt-spun sample of Fe$_2$CoC and measured the MAE for it at 10~K as 0.96~MJ\,m$^{-3}$~\cite{wu2017exploring}.
However, their analysis showed the presence of about 10\% by weight of the bcc Fe phase in the Fe$_2$CoC sample~\cite{wu2017exploring}, which can affect the anisotropy constant.
In contrast, our calculations for about 1/3 Co concentration indicate MAE close to zero.

A feature of our MAE calculations is that they are made for a perfect monocrystal at 0~K, and modeling chemical disorder with VCA leads to overestimation of MAE.
However, the trends thus predicted turned out to be in line with experimental results~\cite{edstrom_magnetic_2015}.
The anisotropy constant of real samples can be significantly affected by numerous additional factors.
In addition to the role of temperature, in the case of (Fe,Co)$_3$C alloys we are faced with the chemical instability of the phase, the presence of microstructure and extraneous phases in the samples, and in the case of nanoparticles, the influence of nanostructuring.
Given the complexity of the problem, the calculated MAEs for monocrystals at 0~K are a good starting point for understanding the relationships governing doping, and the MAE values obtained from the calculations are promising from an application point of view and encourages further exploration of magnetically hard materials in the cementite family.

%-----------spin magnetic moments--------
%
The total spin magnetic moment ($m_s$) is highest for the terminal Fe$_{3}$C phases for both hexagonal and orthorhombic structures, and decrease with Co concentration, see Figs.~\ref{fig:VCA_1D_MAE_ms_ml}(c,d).
We do not observe a pronounced maximum in the intermediate range.
This behavior is similar to that observed, for example, for Fe/Co monoborides~\cite{cadeville_sur_1966,snarski2024searching}.
The observed trend has to do with an increase in the number of electrons per atom with an increase in the concentration of Co in the system.
Additional electrons from Co atoms fill the valence band, especially the 3$d$ orbitals, see Fig.~\ref{fig:DOS}. 
Since the majority Fe~3$d$ spin channels are almost completely occupied, additional electrons fill the minority 3$d$ spin channels, which reduces the occupancy difference between the spin channels, which is what defines the spin magnetic moment.

%-----------orbital magnetic moments--------
%
For (Fe,Co)$_{3}$C alloys, we observe significant differences in the total orbital magnetic moments ($m_l$) determined for various crystallographic directions, see Figs.~\ref{fig:VCA_1D_MAE_ms_ml}(e,f).
Although in some cases orbital moment differences correlate with magnetocrystalline anisotropy, this time we do not observe such relation.
The calculated orbital magnetic moments, see Table~\ref{tab:1D_VCA}, on the Fe sites of 0.04-0.05~$\mu_\mathrm{B}$ are close to the value calculated for bcc Fe of 0.045~$\mu_\mathrm{B}$~\cite{wu1999spin}.
However, both calculated values underestimate the experimental orbital magnetic moment in bcc Fe of 0.086~$\mu_\mathrm{B}$~\cite{chen_experimental_1995}.
The $m_l$ values calculated for Co$_{3}$C phases are lower than the corresponding values for Fe$_{3}$C phases. 
Moreover, the $m_l$ values determined for Co$_{3}$C are lower than the values for fcc Co of 0.071~$\mu_\mathrm{B}$~\cite{budko2019calculation}) and 0.073~$\mu_\mathrm{B}$~\cite{wu1999spin}.

%----------magnetic hardness--------
%
%------hexagonal (Fe,Co)3C---------------
%
Hexagonal (Fe,Co)$_{3}$C alloys with Co concentration from 0 to 30\% are magnetically semi-hard, see Fig.~\ref{fig:VCA_1D_MAE_ms_ml}(g).
For negative values of MAE, seen in the center of the concentration range, we assumed $\kappa$ = 0, see Table~\ref{tab:1D_VCA}.
The hexagonal carbide with 90\%~Co indicates $\kappa$~=~1.13, which defines it as magnetically hard.
%
%------orthorhombic (Fe,Co)3C---------------
%
For orthorhombic alloys, see Fig.~\ref{fig:VCA_1D_MAE_ms_ml}(h), 
semi-hard magnetic properties are observed for most of Co concentrations.
%hard magnetic properties with $\kappa > 1$ for Co contents from 90 to 100\%.

%-----------Curie temperature-------------------------------
%
In the previous section, we discussed the Curie temperature results for the
stoichiometric compounds.
The calculated mean-field Curie temperatures were
1040~K for hexa-Fe$_3$C, 
310~K for hexa-Co$_3$C,
570~K for ortho-Fe$_3$C, and 
458~K for ortho-Co$_3$C.
In Fig.~\ref{fig:tc_ortho_feco3c}, we show the courses of calculated $T_C$ as a function of Co concentration.
For hexa-(Fe,Co)$_3$C phase, the relation is close to linear, while for the ortho-(Fe,Co)$_3$C phase a maximum around 20\% Co is visible.
The close to linear relationship resembles the analogous experimental results for (Fe,Co)B and (Fe,Co)$_2$B alloys~\cite{cadeville_sur_1966, dane_density_2015}, and the maximum in $T_C$ resembles results for bcc FeCo alloys~\cite{sourmail_near_2005}.
As in the discussed trends for MAE and magnetic moment, the trends for $T_C$ depend primarily on the valence band filling of the systems under study increasing with Co concentration, see also Ref.~\cite{lezaic_first-principles_2007}.
As experiments for orhto-Co$_3$C nanoparticles have shown~\cite{mikhalev_magnetic_2019,harris_high_2010,turgut_metastable_2016,el-gendy_enhanced_2014}, we expect that for ortho-(Fe,Co)$_3$C nanoparticles as well, Curie temperatures should be able to be raised to values above 500~K.

%--------section summary------------
%
In summary, we calculated the basic magnetic properties for hexa- and ortho-(Fe,Co)$_3$C alloys.
Hexagonal alloys show a number of promising magnetic properties, such as a high Curie temperature of 1040~K for hexa-Fe$_3$C, and a magnetic hardness ($\kappa$) of 0.83 for hexa-(Fe$_{0.8}$Co$_{0.2}$)$_3$C. 
Nevertheless, their experimental synthesis has so far presented great difficulties, as explained by the positive formation energies of hexa-Fe$_3$C and hexa-Co$_3$C compounds, above those for corresponding orthorhombic phases.
On the other hand, it has been experimentally proven that the ortho-(Fe,Co)$_3$C phase is stabilizable.
DFT calculations predict for it decrease of spin magnetic moment with Co concentration, and maximum in Curie temperature at about 20\% Co.
Experimental Curie temperatures determined for orthorhombic (Fe,Co)$_3$C nanoparticles of more than 500~K offer hope for the practical application of these alloys.
Experiments also showed Curie temperature dispersion for ortho-Co$_3$C nanoparticles, suggesting a strong effect of structuring on the magnetic properties of the system, including magnetic moment, magnetocrystalline anisotropy and Curie temperature.
Most probably due to this feature, the MAE calculated here for ortho-(Fe,Co)$_3$C alloys differ significantly from experimental data.
However, at the same time, calculations and experiments indicate for selected ortho-(Fe,Co)$_3$C compositions the promising MAE values of about 1.0~MJ\,m$^{-3}$ at low temperatures, indicating the possibility of optimizing the alloys to reach similar values above room temperature.
Mentioned optimization would consist of selecting proper Fe/Co ratio, nanoparticle size, and additions of other transition metal elements.
The presence of Co in the alloy is, however, not advantageous from an economic point of view, so alloys with a high Co content are unlikely for applications requiring large amounts of material.

\subsection{Fe$_3$C and Co$_3$C alloys with transition metals}

\begin{figure*}[t]
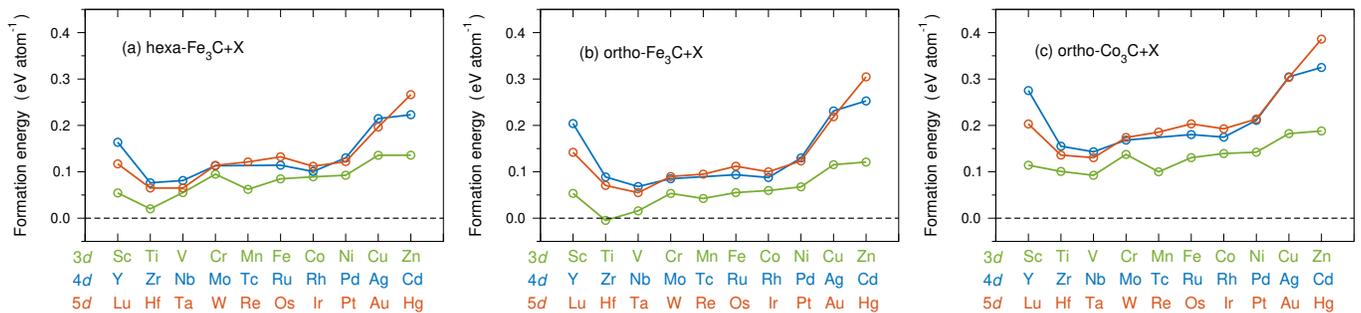

\centering
\includegraphics[trim = 0 0 0 0, clip, height=0.226\textwidth]{hexa-fe3c_Eform_3d_4d_5d.eps}
\includegraphics[trim = -20 0 0 0, clip, height=0.226\textwidth]{ortho-fe3c_Eform_3d_4d_5d.eps}
\includegraphics[trim = -20 0 0 0, clip, height=0.226\textwidth]{ortho-co3c_Eform_3d_4d_5d.eps}
\caption{
\label{fig:E_form}
Formation energies for
(a)~hexa-Fe$_3$C, 
(b)~ortho-Fe$_3$C, and
(c)~ortho-Co$_3$C alloys 
with 3$d$, 4$d$, and 5$d$ transition metals X 
[(Fe/Co)$_{11}$X$_{1}$C$_{4}$].
The calculations were performed with FPLO18 code using the PBE functional and supercell approach. 
}
\end{figure*}

%---------Curie temperature-----------------
%
%
\begin{figure*}[t]
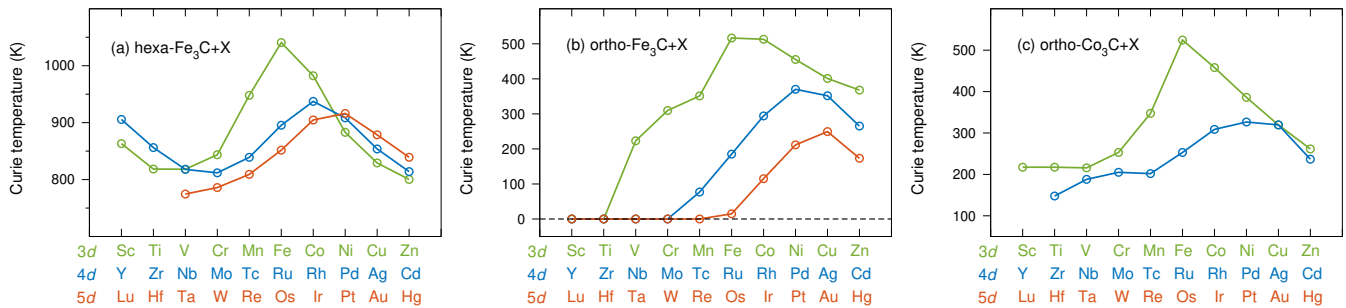

\centering
\includegraphics[trim = 0 0 0 0,clip, height=0.46\columnwidth]{
hexa-fe3c_TC_3d_4d_5d.eps}
\includegraphics[trim = -20 0 0 0,clip, height=0.46\columnwidth]{
ortho-fe3c_TC_3d_4d_5d_0K.eps}
\includegraphics[trim = -20 0 0 0,clip, height=0.46\columnwidth]{figures/ortho-co3c_TC_3d_4d_5d_0K.eps}
\caption{
\label{fig:TC_fe3cx}
The mean-field-theory Curie temperatures ($T_{\mathrm{C}}^{\mathrm{MFT}}$) for hexa- and ortho-(Fe$_{0.916}$X$_{0.084}$)$_3$C alloys with 3$d$, 4$d$, and 5$d$ transition metals X.
The results were calculated using the FPLO5 code and the Perdew and Wang (PW92) functional.
The chemical disorder was modeled with coherent potential approximation (CPA) and the paramagnetic state with disorder local moment (DLM-CPA) method.
%
%For ortho-Fe$_3$C and ortho-Co$_3$C phases, the experimental Curie temperatures are 481~K~\cite{nicholson_solubility_1957} and about 650~K~\cite{el-gendy_enhanced_2014}, respectively.
}
\end{figure*}

\begin{figure*}[t]
\centering
\includegraphics[clip, height=0.23\textwidth]{Fe3C_Hexa_MS_3d_4d_5d.eps}
\hfill
\includegraphics[clip, height=0.23\textwidth]{Fe3C_Hexa_ms_X_3d_4d_5d.eps}
\hfill
\includegraphics[clip, height=0.23\textwidth]{Fe3C_Hexa_Ml_3d_4d_5d.eps}\\
\vspace{4mm}
\includegraphics[clip, height=0.23\textwidth]{Fe3C_Hexa_MAE_3d_4d_5d.eps}
\hfill
\includegraphics[clip, height=0.23\textwidth]{Fe3C_Hexa_kappa_3d_4d_5d.eps}
\hfill
\includegraphics[clip, height=0.23\textwidth]{Fe3C_Hexa_DE32.eps}
\caption{
\label{fig:Hexa_doped}
The magnetic properties of hexa-Fe$_3$C alloys with 3$d$, 4$d$, and 5$d$ transition metals X (Fe$_{11}$X$_{1}$C$_{4}$):
(a)~total spin magnetic moment (per formula unit);
(b, c)~spin and orbital magnetic moments on substitution atoms X (per TM atom);
(d)~magnetocrystalline anisotropy energy (MAE);
(e)~magnetic hardness; and 
(f)~energy difference (DE$_{32}$) between the two higher energies among the three total energies calculated along the three main crystallographic axes.
The calculations were performed with FPLO18 code using the PBE functional and supercell approach. 
%
%Full circles indicate results for the base material (hexa-Fe$_3$C) without dopant of other type.
}
\end{figure*}

%----ortho-Fe3C+X-----
%
\begin{figure*}[t]
\centering
\includegraphics[clip, height=0.23\textwidth]{Fe3C_Cementyt_MS_3d_4d_5d.eps}
\hfill
\includegraphics[clip, height=0.23\textwidth]{Fe3C_Cementyt_ms_X_3d_4d_5d.eps}
\hfill
\includegraphics[clip, height=0.23\textwidth]{Fe3C_Cementyt_Ml_3d_4d_5d.eps}\\
\vspace{4mm}
\includegraphics[clip, height=0.23\textwidth]{Fe3C_Cementyt_MAE_3d_4d_5d.eps}
\hfill
\includegraphics[clip, height=0.23\textwidth]{Fe3C_Cementyt_kappa_3d_4d_5d.eps}
\hfill
\includegraphics[clip, height=0.23\textwidth]{Fe3C_Cementyt_DE32.eps}
\caption{
\label{fig:Cementyt_doped}
The magnetic properties of ortho-Fe$_3$C alloys with 3$d$, 4$d$, and 5$d$ transition metals X (Fe$_{11}$X$_{1}$C$_{4}$):
(a)~total spin magnetic moment (per formula unit);
(b, c)~spin and orbital magnetic moments on substitution atoms X (per TM atom);
(d)~magnetocrystalline anisotropy energy (MAE);
(e)~magnetic hardness; and 
(f)~energy difference (DE$_{32}$) between the two higher energies among the three total energies calculated along the three main crystallographic axes.
The calculations were performed with FPLO18 code using the PBE functional and supercell approach. 
}
\end{figure*}

\begin{figure*}[t]
\centering
\includegraphics[clip, height=0.23\textwidth]{Co3C_Cementyt_MS_3d_4d_5d.eps}
\hfill
\includegraphics[clip, height=0.23\textwidth]{Co3C_Cementyt_ms_X_3d_4d_5d.eps}
\hfill
\includegraphics[clip, height=0.23\textwidth]{Co3C_Cementyt_Ml_3d_4d_5d.eps}\\
\vspace{4mm}
\includegraphics[clip, height=0.23\textwidth]{Co3C_Cementyt_MAE_3d_4d_5d.eps}
\hfill
\includegraphics[clip, height=0.23\textwidth]{Co3C_Cementyt_kappa_3d_4d_5d.eps}
\hfill
\includegraphics[clip, height=0.23\textwidth]{Co3C_Cementyt_DE32.eps}
\caption{
The magnetic properties of ortho-Co$_3$C alloys with 3$d$, 4$d$, and 5$d$ transition metals X (Co$_{11}$X$_{1}$C$_{4}$):
(a)~total spin magnetic moment (per formula unit of two atoms);
(b, c)~spin and orbital magnetic moments on substitution atoms X (per TM atom);
(d) magnetocrystalline anisotropy energy (MAE);
(e) magnetic hardness; and 
(f) energy difference (DE$_{32}$) between the two higher energies among the three total energies calculated along the three main crystallographic axes.
The calculations were performed with FPLO18 code using the PBE functional and supercell approach. 
%
%Full circles indicate results for the base material (ortho-Fe$_3$C) without dopant of other type.
}
\label{fig:Co_doped}
\end{figure*}

%-------introduction-------------------
%
In this section, we will discuss calculation results for Fe$_3$C and Co$_3$C alloys with transition metals.
In our models, the transition metal atom replaces one of the twelve Fe (or Co) atoms in the Fe$_{11}$X$_{1}$C$_{4}$ (or Co$_{11}$X$_{1}$C$_{4}$) supercell, see Fig.~\ref{tab:struct}.
Similar models of ortho-Fe$_{11}$X$_{1}$C$_{4}$ alloys have been investigated by Shein~\textit{et al.}~\cite{shein_electronic_2007}.
Below we will present the results of calculations of formation energies, magnetic moments, and magnetocrystalline anisotropy energies.
These results are further complemented by Curie temperature calculations performed using the disorder local moments (DLM) method.
In the following analysis, we consider hexa-Fe$_3$C, orhto-Fe$_3$C, and ortho-Co$_3$C alloys, ignoring hexa-Co$_3$C alloys.
The decision to exclude hexa-Co$_3$C alloys is due to the unpromising properties of the hexa-Co$_3$C compound (magnetic hardness equal zero and formation energy higher than for ortho-Co$_3$C phase).

\subsubsection{Formation energy}
%
%-------formation energies-------------------
%
Figure~\ref{fig:E_form} presents the calculated formation energies for 
hexa-Fe$_3$C, 
ortho-Fe$_3$C, and 
ortho-Co$_3$C alloys with transition metals.
With one exception (ortho-Fe$_{11}$Ti$_{1}$C$_{4}$), all alloys have positive formation energies, implying lack of chemical stability.
The positive values are in most cases relatively small, making it possible to stabilize the materials by forming nanoparticles, alloys, or composites~\cite{nicholson_solubility_1957, umemoto_influence_2001, konyaeva_electronic_2009, wu2017exploring}.

%------hexa-Fe3C+X----------
%
%The calculated formation energy of hexa-Fe$_3$C is about 0.085~eV\,atom$^{-1}$ and is slightly higher than that for ortho-Fe$_3$C (0.055~eV\,atom$^{-1}$).
%
%The situation is similar when comparing trends for hexa-Fe$_3$C and ortho-Fe$_3$C alloys with transition metals.
%
%------ortho-Fe3C+X----------
%
%The calculated formation energy of ortho-Fe$_3$C (0.055~eV\,atom$^{-1}$) stays in a good agreement with previous theoretical and experimental results (0.05--0.08~eV\,atom$^{-1}$)~\cite{shein_electronic_2007, konyaeva_electronic_2009, ande2012first}.

%
The plots of dependencies of formation energies on 3$d$ and 4$d$ transition metal substitutions in ortho-Fe$_3$C are almost identical to the previously mentioned theoretical results~\cite{shein_electronic_2007, ande2012first}.
Although the formation energies of the alloys studied provide valuable information, the idealized single-phase nature of the models means that the actual impact of additives is best verified experimentally.
Bulk ortho-Fe$_3$C has been tested by Umemoto~\textit{et al.}~\cite{umemoto_influence_2001}, who found that Cr, Mn, V and Mo additions stabilize the cementite, whereas Ti, Ni, and Si destabilize it.
Furthermore, it has been determined that the substitution of 5~at.\% Ti leads to the appearance of additional phases of TiC and $\alpha$-Fe~\cite{umemoto_influence_2001}.
%
%---------ortho-Co3C+X----------
%
The formation energies of ortho-Co$_3$C alloys are higher than those of ortho-Fe$_3$C, however, still relatively low.

\subsubsection{Curie temperature}
%
%---------Curie temperature-----------------
%

%--------DLM TC results--------------
%
In Fig.~\ref{fig:TC_fe3cx} we present the mean-field theory Curie temperatures ($T_{\mathrm{C}}^{\mathrm{MFT}}$) calculated for hexa-(Fe$_{0.916}$X$_{0.084}$)$_3$C, ortho-(Fe$_{0.916}$X$_{0.084}$)$_3$C,
and ortho-(Co$_{0.916}$X$_{0.084}$)$_3$C 
alloys with transition metals X.
The transition metal concentration of 0.084 was chosen to reproduce the $\frac{1}{12}$ concentration of transition metal element X in Fe$_{11}$X$_1$C$_4$ models.

%---------hexa-Fe3C+X---------
%
In Fig.~\ref{fig:TC_fe3cx} we see, that the dependencies of $T_C$ on the atomic number of substitution have a clear sinusoidal trend and
that each substitution reduces the $T_C$ of the parental hexa-Fe$_3$C compound.
Nevertheless, all the considered hexa-Fe$_3$C~+~X alloys show relatively high $T_C$.

%---------ortho-Fe3C+X---------
%
Calculated trends for transition metal elements suggest that all substitutions, except Co, will significantly reduce the $T_C$ of the alloy.
Whereas, the elements 4$d$ and 5$d$ will have stronger effect than the elements 3$d$.
The transition metal elements on the left side of the periods reduce $T_C$ to zero, which means the absence of magnetic ordering.
Previous experiments confirm the reduction of Curie temperature after alloying of cementite with Cr, Mn, or Mo~\cite{kagawa_lattice_1979, schaaf_mossbauer_1992}.
However, it has been measured that up to about 2\%~Ni raises the Curie temperature of the cementite~\cite{kagawa_lattice_1979}.
The effect of lowering $T_C$ can make it difficult or impossible to use most transition metals as additives to affect MAE and magnetic hardness.
The solution could be, for example, a combination of transition metal alloying on the Fe site with B alloying on the C side.
Indeed, the experiment showed a linear growth in $T_C$ up to 851~K with increasing B concentration in ortho-Fe$_3$(B,C) borocarbides~\cite{nicholson_solubility_1957}.

%---------ortho-Co3C+X---------
%
Calculations indicate that only Fe substitution raises the $T_C$ of the ortho-Co$_3$C-based system, while all other elements considered lower it.
The determined $T_C$ values below and well below 400~K suggest the impossibility of using most ortho-Co$_3$C~+~X alloys as permanent magnets.
The solution, as we suggested above, could be the alloying of B at the C position, since, as shown experimentally, the $T_C$ of ortho-Co$_3$B is much higher than the value for ortho-Co$_3$C and is~750~K~\cite{pal_properties_2017}.

\subsubsection{Magnetic moments and magnetic hardness}

%------introduction------------
%
After discussing the formation energy and Curie temperature, we will show how alloying with transition metals affect the magnetic moments and magnetic hardness, see Figs.~\ref{fig:Hexa_doped}, \ref{fig:Cementyt_doped}, and \ref{fig:Co_doped} for hexa-Fe$_{11}$X$_{1}$C$_{4}$, ortho-Fe$_{11}$X$_{1}$C$_{4}$, and ortho-Co$_{11}$X$_{1}$C$_{4}$ alloys, respectively.
In the majority of instances, clear patterns emerge with regard to the dependencies.
Trends in magnetic moments are similar to those calculated previously for alloys of ortho-Fe$_{3}$C~\cite{shein_electronic_2007}, CeFe$_{12}$~\cite{snarski2022effect}, monoborides~\cite{snarski2024searching}, and Fe~\cite{akai_nuclear_1988,dederichs_ab-initio_1991,snarski-adamski_magnetic_2022}.
MAE trends, on the other hand, do not resemble previous findings.
\vspace{2mm}

\paragraph{Hexa-Fe$_{3}$C alloys with transition metals \vspace{3mm}}
%
%---------magnetic moments------------
%
We will begin our discussion of the results with hexa-Fe$_{3}$C alloys.
Figure~\ref{fig:Hexa_doped} presents a collection of the magnetic properties for alloys with 3$d$, 4$d$ and 5$d$ metals.
The spin magnetic moment trends for the three series of transition metals have a waveform shape, see Figs.~\ref{fig:Hexa_doped}(a,b).
The parent system without substitution, hexa-Fe$_{3}$C, shows the highest total spin magnetic moment.
All transition metal substitutions are spin polarized in the ferromagnetic medium and contribute to the total magnetic moment of the alloy, lowering its value.
The orbital magnetic moments on the substitution sites are much smaller from spin magnetic moments, and oscillate between -0.07 and +0.06~$\mu_\mathrm{B}$\,atom$^{-1}$, see Fig.~\ref{fig:Hexa_doped}(c).

%---------MAE and kappa----------
%
Figures~\ref{fig:Hexa_doped}(d) and \ref{fig:Hexa_doped}(f) show the magnetocrystalline anisotropy energies (MAE and DE$_{32}$) for hexa-Fe$_{11}$X$_{1}$C$_{4}$ alloys. 
From all considered compositions, the system with Pt has the highest MAE of 1.73~MJ\,m$^{-3}$. 
This value corresponds to the highest magnetic hardness of 0.96, see Fig.~\ref{fig:Hexa_doped}(e). 
Our previous results for transition metal alloying in the Fe matrix, also pointed to Pt as a metal that stands out for its exceptionally positive effect on MAE~\cite{snarski-adamski_magnetic_2022}.
Unfortunately, the high price of platinum, is not in favor of potential applications.

\vspace{2mm}

\paragraph{Ortho-Fe$_{3}$C alloys with transition metals \vspace{2mm}}
%
%
%---------magnetic moments------------
%
Trends of magnetic moments for ortho-Fe$_3$C alloys with transition metal elements, see Fig.~\ref{fig:Cementyt_doped}(a-c), are similar to those presented above for hexa-Fe$_3$C alloys.
However, the values of total spin magnetic moments are lower by about 0.5~$\mu\mathrm{_B}$\,f.u.$^{-1}$.
Similar trends in local magnetic moments on transition metal substitutions were also observed in previous calculations for ortho-Fe$_{11}$X$_{1}$C$_{4}$ alloys with 3$d$ and 4$d$ elements~\cite{shein_electronic_2007}.
Among the ortho-Fe$_{11}$X$_{1}$C$_{4}$ alloys, the parent ortho-Fe$_3$C compound has the highest total spin magnetic moment of 5.60~$\mu\mathrm{_B}$\,f.u.$^{-1}$.

%---------MAE and kappa----------
%
The MAE and magnetic hardness of ortho-Fe$_3$C alloys appear more promising than that of hexa-Fe$_3$C alloys, see Fig.~\ref{fig:Cementyt_doped}(d,e). 
The MAE increases above 1~MJ\,m$^{-3}$ for Re, Ir, and Pt substitutions, leading to MAE values of 1.4, 1.2, and 1.1~MJ\,m$^{-3}$, respectively.
However, as shown in Fig.~\ref{fig:Cementyt_doped}(e), none of the alloys can be classified as a hard magnetic ($\kappa > 1$). 
The magnetic hardness closest to exceeding unity (0.93) has an alloy with Re.
In addition, as we showed earlier, alloying ortho-Fe$_3$C with 5$d$ transition metals leads to a strong decrease in Curie temperature.
It makes the system with Re ultimately not a good candidate for a permanent magnet either.

%-----DE32--------
%
In systems (such as orthorhombic) that do not have a unique distinguished crystallographic axis, we must be careful when considering MAE values, checking DE$_{32}$, see Sec.~\ref{Calculation details} for definition.
The DE$_{32}$ equal to 0.8~MJ\,m$^{-3}$ for ortho-Fe$_{11}$Re$_{1}$C$_{4}$ with MAE of 1.4~MJ\,m$^{-3}$ suggests uniaxial magnetocrystalline anisotropy.
On the other hand, DE$_{32}$ of around 2.4~MJ\,m$^{-3}$ for ortho-Fe$_{11}$Os$_{1}$C$_{4}$, compared to MAE of 0.4~MJ\,m$^{-3}$ suggests magnetocrystalline anisotropy close to easy-plane.
\vspace{2mm}

\paragraph{Ortho-Co$_{3}$C alloys with transition metals \vspace{3mm} \newline}

%-------motivation-------
%
The magnetic hardness of the ortho-Co$_3$C is one of the highest among the considered orthorhombic compositions, see Fig.~\ref{fig:VCA_1D_MAE_ms_ml}.
%
%------magnetic moments-------
%
The magnetic properties calculated for ortho-Co$_3$C alloys are shown in Fig.~\ref{fig:Co_doped}. 
For the total spin magnetic moment, see Fig.~\ref{fig:Co_doped}(a),
we observe sinusoidal behavior, like before for hexa-Fe$_3$C alloys.
However, the total magnetic moments calculated for ortho-Co$_3$C alloys are about half those for Fe$_3$C alloys.
The maximum of spin magnetic moment, for Co$_{11}$Mn$_{1}$C$_{4}$, is equal to 3.4~$\mu_\mathrm{B}$~per~f.u. and the minimum, for Co$_{11}$W$_{1}$C$_{4}$, is equal to 2.2~$\mu_\mathrm{B}$~per~f.u.
Only for Mn and Fe substitutions, the spin magnetic moment is higher than for Co in the initial ortho-Co$_3$C phase, see Fig.~\ref{fig:Co_doped}(b).
The orbital magnetic moments on the substituted sites, ranging from -0.04 to +0.04~$\mu_\mathrm{B}$\,atom$^{-1}$, are also about twice smaller than in the case of Fe$_3$C alloys, see Fig.~\ref{fig:Co_doped}(c).

%----------MAE and kappa-------------
%
In Figure~\ref{fig:Co_doped}(d), we see once again that the MAE significantly depends on the substituting element.
The calculated MAE results take a range from zero to 0.8~MJ\,m$^{-3}$.
The MAE and magnetic hardness calculations indicate that Co$_{11}$W$_1$C$_4$ and Co$_{11}$Os$_1$C$_4$ can be classified as hard magnetic ($\kappa > 1$) with MAE values equal to 0.5 and 0.8~MJ\,m$^{-3}$, respectively, see Figs.~\ref{fig:Co_doped}(d,e).
Unfortunately, as for ortho-Fe$_3$C~+~X alloys, 5$d$ substitutions will strongly lower the Curie temperature, making their use impractical.
%
%-------Co3C summary-------
%
Alloying of ortho-Co$_3$C with transition metals in most cases causes a decrease in magnetic hardness.
The small increase in magnetic hardness predicted for alloying with W and Os is burdened by the high prices of these elements and expected significant decrease in Curie temperature.

%-------section summary----
%
In summary, calculations of Curie temperature and magnetic hardness do not suggest that alloying with transition metals of hexa-Fe$_3$C, ortho-Fe$_3$C, and ortho-Co$_3$C compounds lead to a significant improvement in their potential for permanent magnet applications.
The magnetic hardness of hexa-Fe$_3$C~+~X alloys, except for Pt substitution, does not exceed 0.7, and in addition, the formation energies of hexagonal alloys are higher than those of orthorhombic alloys, making them difficult to obtain.
In the case of ortho-Fe$_3$C~+~X alloys, Re, Ir, and Pt lead to an increase in magnetic hardness but at the same time lower the Curie temperature to values far below room temperature, which disqualifies potential applications for permanent magnets.
In the case of the most promising hard magnetic ortho-Co$_3$C compound, a few 5$d$ transition metals (Re, Pt, Au) do lead to a slight increase in magnetic hardness, but at the price of a significant drop in Curie temperature.

\subsection{(Fe,Co)$_3$(B,C,N) alloys}

\begin{figure*}[t]
\centering
\includegraphics[clip, width=0.29\linewidth]{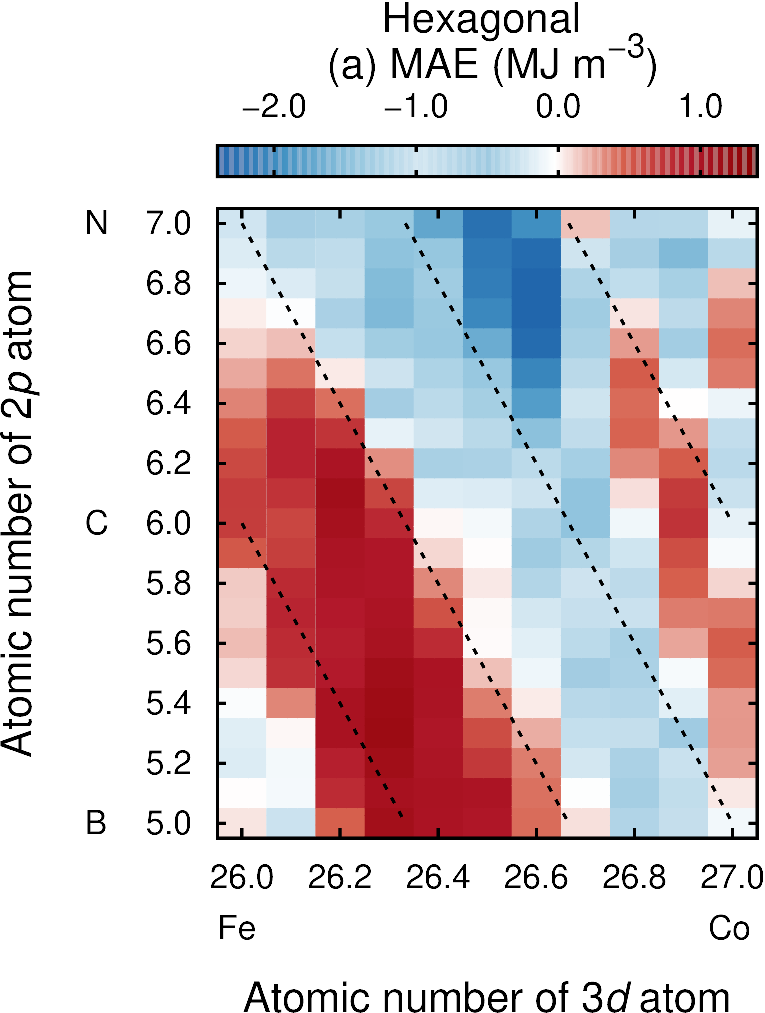}
\hfill
\includegraphics[clip, width=0.22\linewidth]{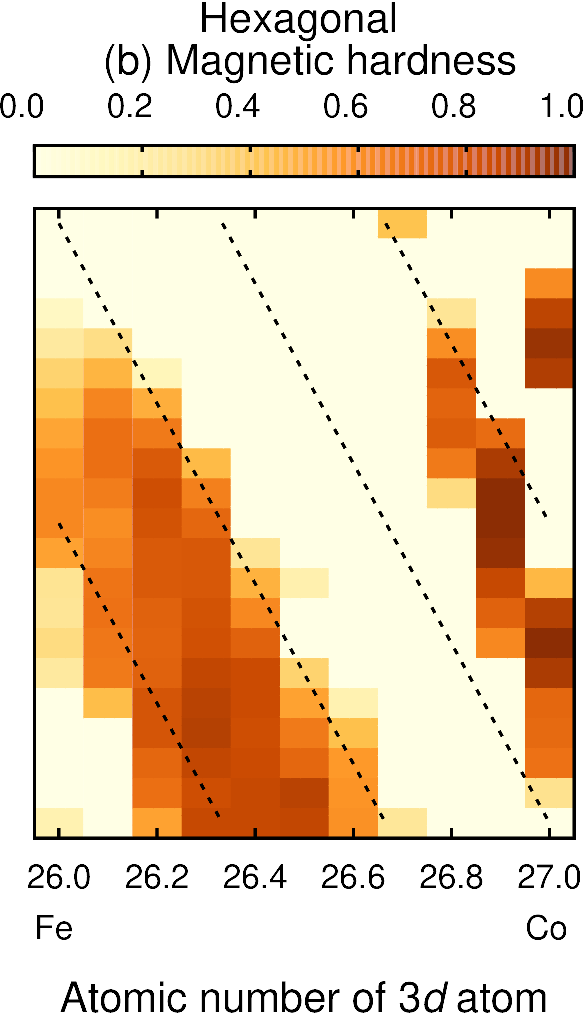}
\hfill
\includegraphics[clip, width=0.22\linewidth]{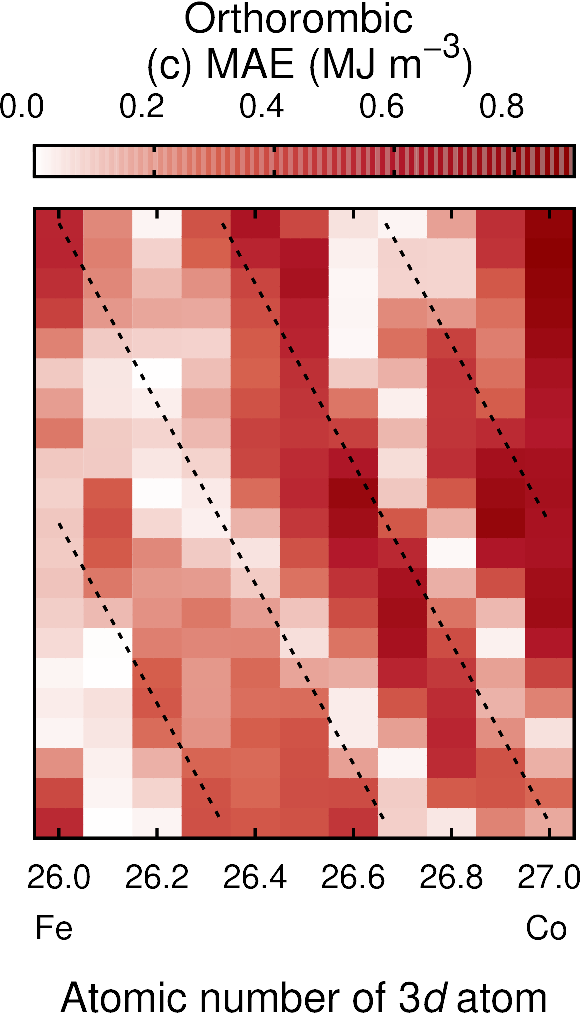}
\hfill
\includegraphics[clip, width=0.22\linewidth]{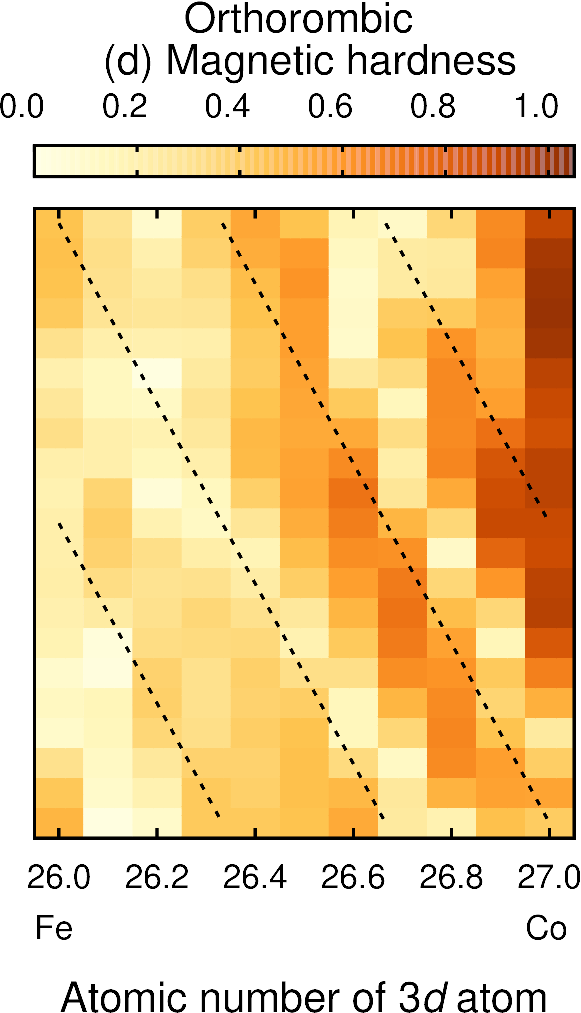}
\caption{
(a,c) Magnetocrystalline anisotropy energy and (b,d) magnetic hardness as dependencies on the average atomic number of the elements 3$d$ and 2$p$ calculated for hexagonal and orthorhombic phase of (Fe,Co)$_{3}$(B,C,N) pseudobinary alloys.
Calculations performed with virtual crystal approximation (VCA), using the FPLO18 code with PBE exchange-correlation potential.
Along the dashed isolines, the total atomic number per formula (and thus the total number of electrons per formula) is constant.
\label{fig:2D}
}
\end{figure*}
\begin{figure}[t]
\centering
\includegraphics[trim = 15 210 30 100,clip, width=0.95\columnwidth]{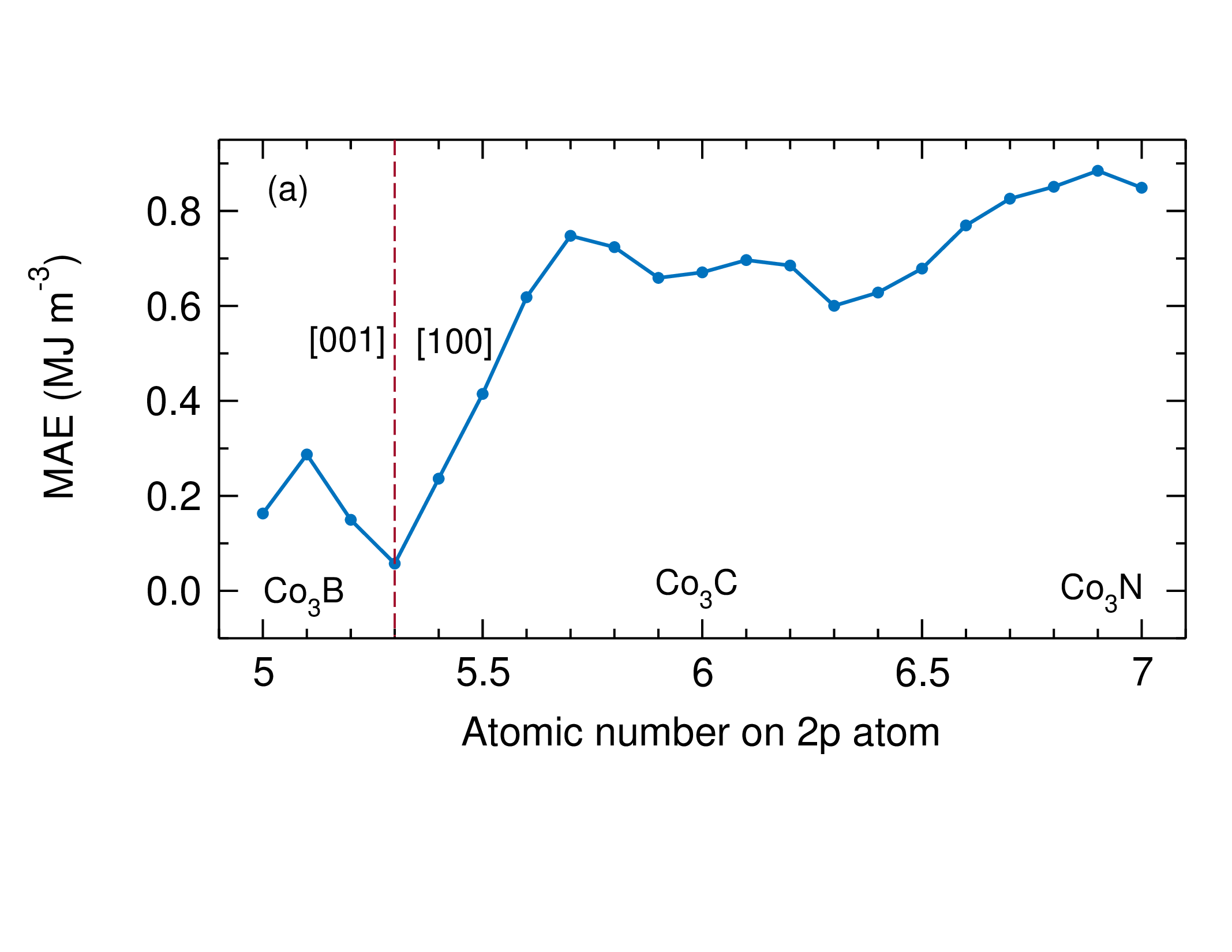}
\includegraphics[trim = 15 210 30 100, clip, width=0.95\columnwidth]{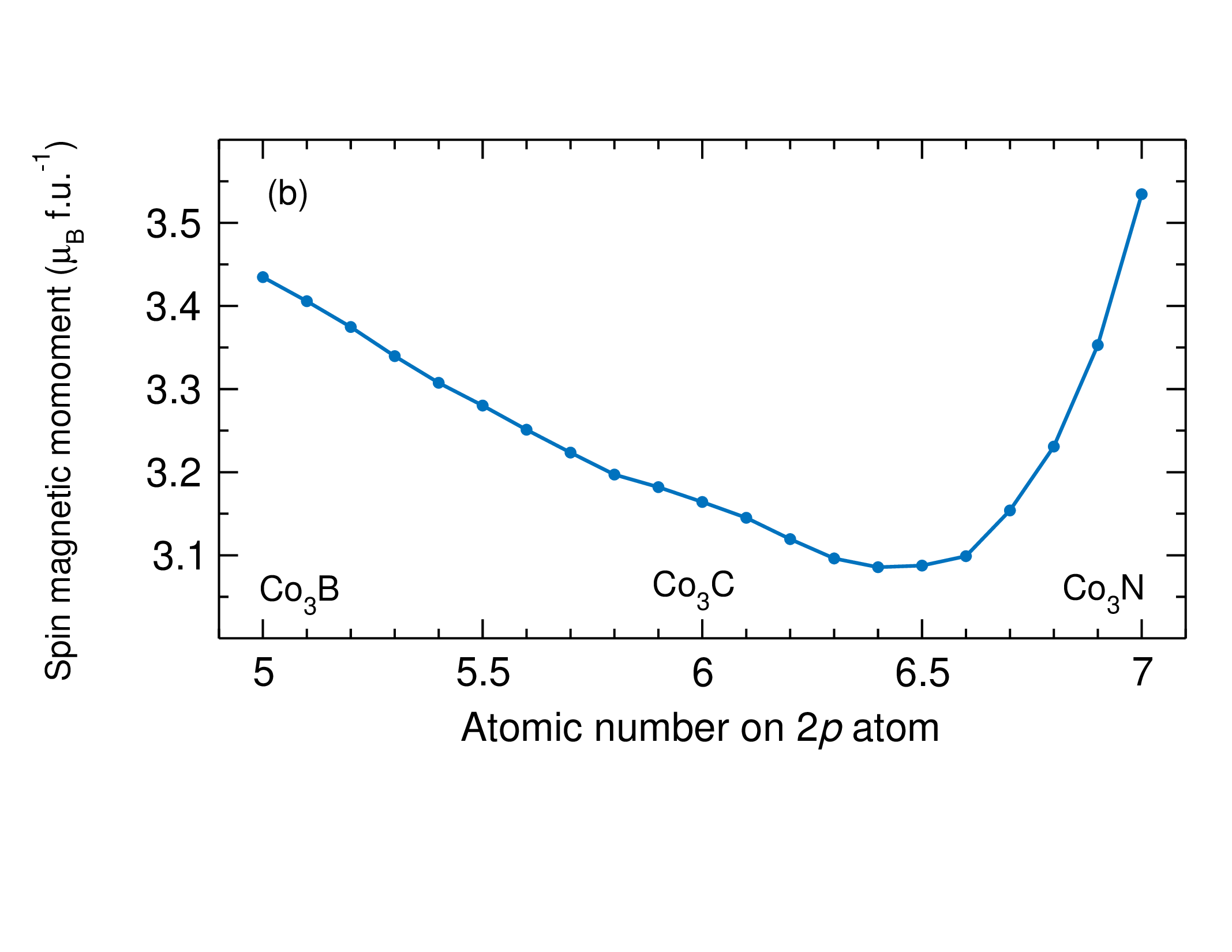}
\includegraphics[trim = 15 150 30 100, clip, width=0.95\columnwidth]{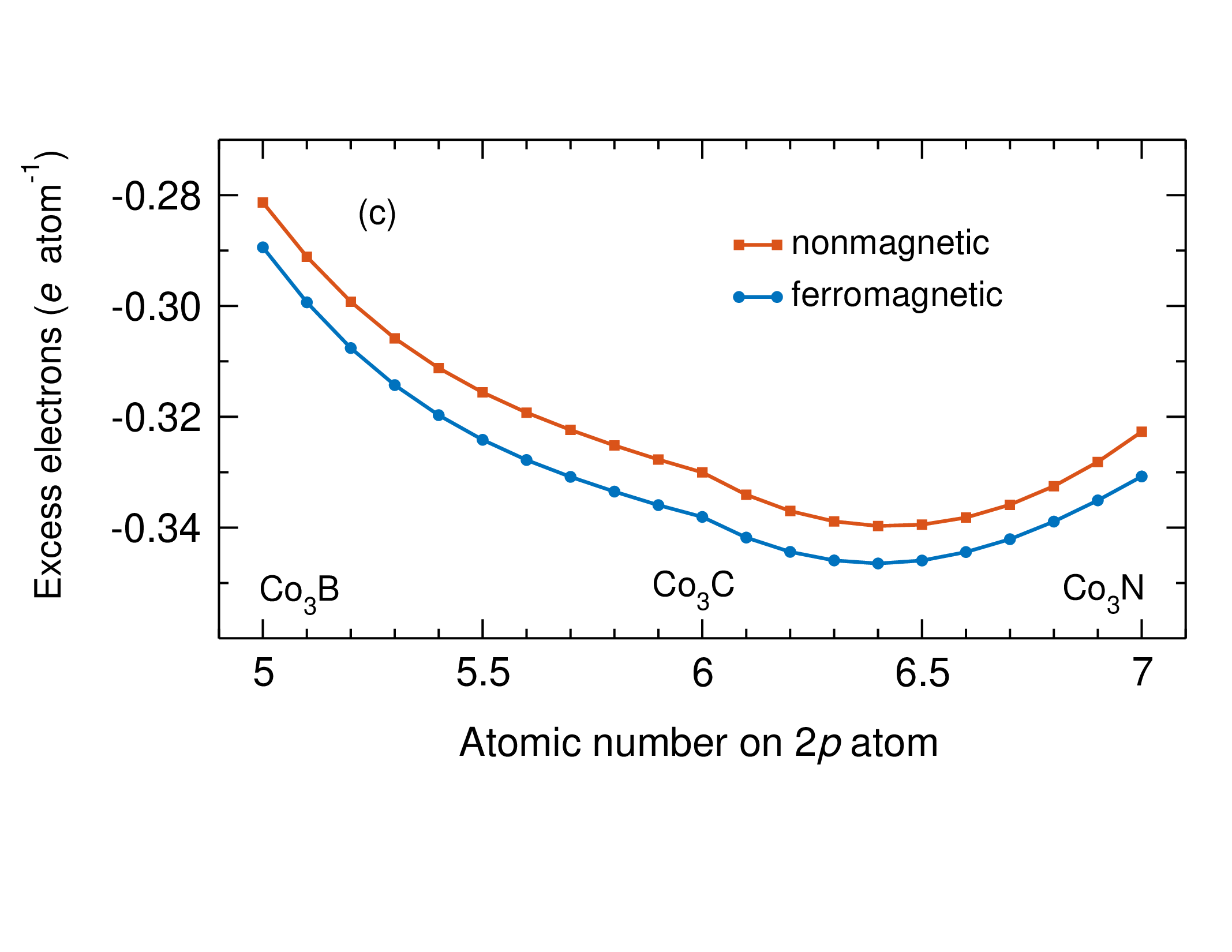}
\caption{
\label{fig:Co3(B,C,N)dep}
The Magnetocrystalline anisotropy energy (a), total spin magnetic moment (b), and excess electrons at the Co site (c) as a function of
the average atomic number of elements 2$p$ calculated for orthorhombic Co$_3$(B,C,N) alloys.
In the MAE plot, the red dashed line separates the regions with different directions of magnetic easy axis. 
Calculations were performed with virtual crystal approximation (VCA), using the FPLO18 code with PBE exchange-correlation potential.
}
\end{figure}

\begin{figure}[t]
\centering
\includegraphics[trim = 30 150 30 40,clip, width=0.9\columnwidth]{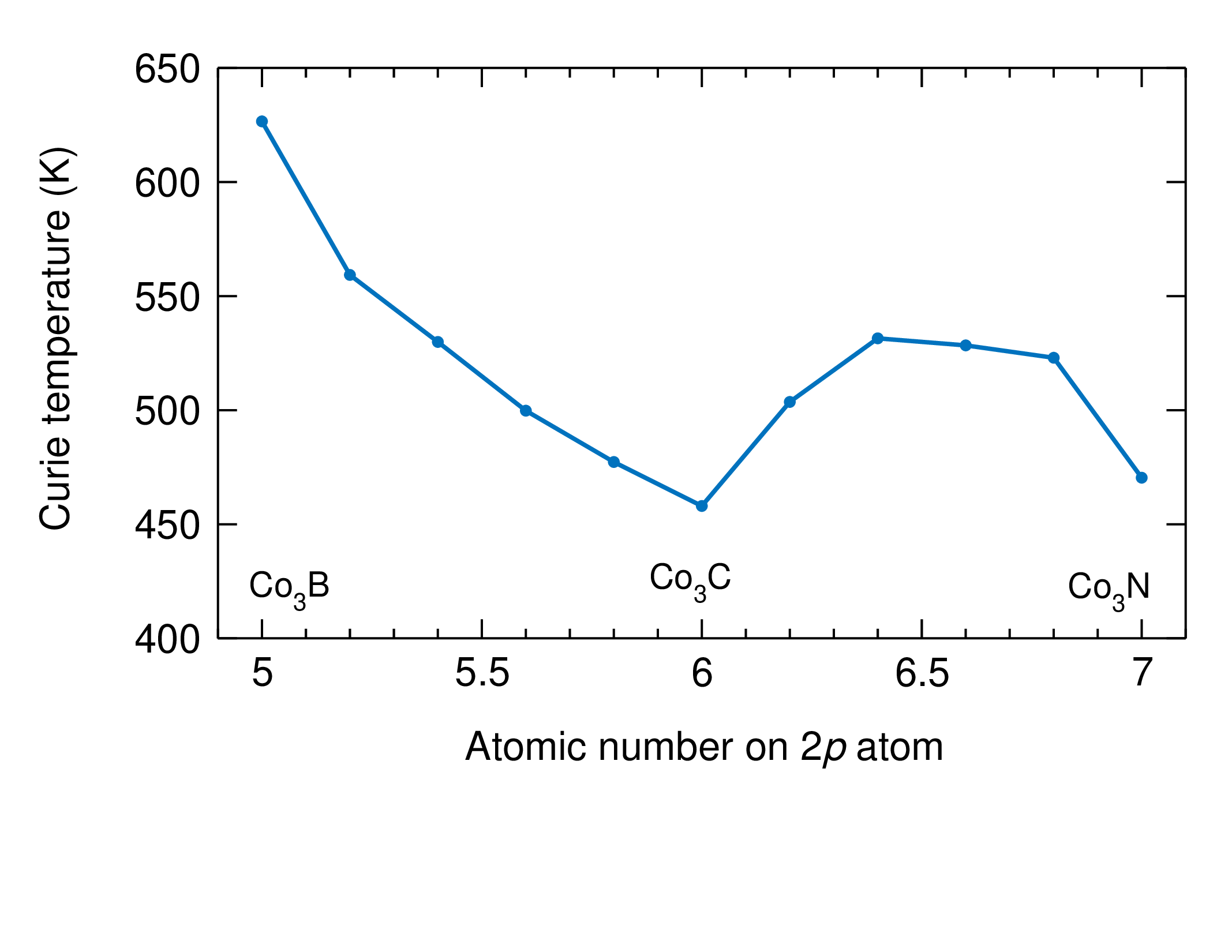}
\caption{
\label{fig:tc_co3bcn}
The mean-field-theory Curie temperatures ($T_{\mathrm{C}}^{\mathrm{MFT}}$) calculated for orthorhombic Co$_3$(B,C,N) alloys as a function of the average atomic number of 2$p$ element.
The results were calculated using the FPLO5 code and the Perdew and Wang (PW92) functional.
The chemical disorder was modeled with coherent potential approximation (CPA) and the paramagnetic state with disorder local moment (DLM-CPA) method.
}
\end{figure}

\begin{table}[t]
\caption{\label{tab:2DMap_values} 
Magnetocrystalline anisotropy energy [MAE (MJ\,m$^{-3}$)] and
magnetic hardness [$\kappa$] of selected (Fe,Co)$_{3}$(B,C,N) compositions with hexagonal and orthorhombic symmetries
calculated with FPLO18 code using PBE functional and virtual crystal approximation to model alloying. 
}
\begin{ruledtabular}
\begin{tabular}{ccccc}		
Compound/Alloy	&\multicolumn{2}{c}{Hexagonal}  & \multicolumn{2}{c}{Orthorhombic}\\
& MAE &   $\kappa$  & MAE  & $\kappa$     \\
\hline													
Fe$_{3}$B			                                       &	0.05  & 0.18 &  0.53  & 0.50 \\
Fe$_{3}$C		                                           &	0.68  & 0.63 &  0.11  & 0.21  \\
Fe$_{3}$N	                                               &	-0.25 & 0   &  0.55  &	0.45   \\
Co$_{3}$B                                                  &	-0.09 & 0	 &  0.16  &	0.41    \\
Co$_{3}$C                                                  &	-0.15 & 0	 &  0.67  &	0.91     \\
Co$_{3}$N                                                  &	-0.15 & 0   & 0.82	  &	0.92      \\
\hline
Co$_{3}$(C$_{0.3}$N$_{0.7}$)                               & --       & --   & 0.83   & 1.02  	   \\
Co$_{3}$(B$_{0.4}$C$_{0.6}$)                                 & 0.47     & 1.01 &  --    &	-- 	   \\
(Fe$_{0.1}$Co$_{0.9}$)$_{3}$C                              & 0.75     & 1.13 &  --    &	-- 	   \\
(Fe$_{0.1}$Co$_{0.9}$)$_{3}$(C$_{0.9}$N$_{0.1}$)            & 0.69     & 1.10 &  --    &	-- 	   \\
(Fe$_{0.7}$Co$_{0.3}$)$_{3}$(B$_{0.7}$C$_{0.3}$)             &	1.19  &	0.91 &  --    &	-- 	   \\
\end{tabular}
\end{ruledtabular}
\end{table}

%-------introduction of the section; borocarbides and boronitrides ---------
%
In Sec.~\ref{sec:feco3c} we discussed (Fe/Co)$_3$C alloys in which substituting takes place at the 3$d$ sites (Fe and Co).
An additional degree of freedom in controlling the composition of Fe$_3$C-type alloys is substituting at the 2$p$ (carbon) site~\cite{nicholson_solubility_1957, du_reevaluation_1993_1,medvedeva2007simulation}.
Two natural candidates for such alloying are boron and nitrogen -- carbon's neighbors in the periodic table.
Although Fe$_3$(B,C) borocarbides have been synthesized since at least the 1950s~\cite{nicholson_solubility_1957}, they have not received wider attention.
However, replacing carbon with boron in Fe$_3$(B,C) alloys leads to significant changes in properties.
Nicholson showed experimentally that in the alloys range from Fe$_3$C to Fe$_3$B, the magnetic moment grows by about 10\% and the Curie temperature linearly increases from 481~K to the impressive 824~K~\cite{nicholson_solubility_1957}.
A similar significant increase in Curie temperature during alloying at the $p$-site was also observed for the transition from Fe$_5$PB$_2$ to Fe$_5$SiB$_2$, where an alloying element with one less atomic number was also used~\cite{hedlund_magnetic_2017}.
Medvedeva~\textit{et al.} have shown by supercell method that B alloying increases the magnetic moment of ortho-Fe$_3$(B,C) borocarbides and stabilizes them~\cite{medvedeva2007simulation}.
Ande and Sluiter, also using supercell calculations, tested the effect of substitution of Al, Si, P, or S at the C-site in ortho-Fe$_3$C on the stability of solid solutions~\cite{ande2012first}.
Moreover, the hexa-Fe$_3$N nitride has a negative formation energy, hence it stabilizes the hexagonal nitrocarbides~\cite{rounaghi_synthesis_2019}.
The hexa-Fe$_3$(C,N) powder consisting of 5\% C was previously synthesized~\cite{rounaghi_synthesis_2019} and classified as magnetically soft.
%
%
%

%-------introduction to 2d map results-----------
%
In this section, we extend the study of (Fe,Co)$_3$C alloys, and analyze the alloying at both the 3$d$ and 2$p$ sites within a single model.
We consider (Fe,Co)$_{3}$(B,C,N) alloys
with concentration range at 3$d$ site from Fe to Co (atomic numbers from 26 to 27) and at 2$p$ site from B, through C, to N (atomic numbers from 5 to 7).
In Figure~\ref{fig:2D}, we show maps of MAE and magnetic hardness for hexagonal and orthorhombic alloys.
Furthermore, in Table~\ref{tab:2DMap_values}, we present calculated values of MAE and magnetic hardness for stoichiometric compounds and for alloys with the highest determined magnetic hardnesses.

%-------MAE vs number of electrons - discussion-------
%
Studying alloying simultaneously at 3$d$ and 2$p$ sites allows conclusions to be drawn about the basic nature of the MAE.
Particularly noticeable in MAE maps are diagonal stripes, see Fig.~\ref{fig:2D}.
It means that the MAE does not depend on the details of the composition, but on the total atomic number of the system (or equivalent total number of electrons).
The dependence of the MAE on the number of electrons is in agreement with the model of the origin of the MAE as a subtle property, which depends on the details of the valence band form and its filling level~\cite{edstrom_magnetic_2015}.

%----------hexa phase - 2d map----------------
%
The hexagonal (Fe,Co)$_{3}$(B,C,N) system exhibit MAE from -2.4 to 1.19~MJ\,m$^{-3}$, see Fig.~\ref{fig:2D}.
The highest value of MAE (1.19~MJ\,m$^{-3}$) is obtained for (Fe$_{0.7}$Co$_{0.3}$)$_{3}$(B$_{0.7}$C$_{0.3}$).
Several other compositions also exhibit magnetic hardness above unity, see Table~\ref{tab:2DMap_values}. 
%
%--------ortho phase - 2d map---------------
%
Moreover, the orthorhombic (Fe,Co)$_{3}$(B,C,N) system shows only positive MAE values, while the direction of the magnetic easy axis changes.
For orthorhombic phases, the highest values of magnetic hardness are observed for full Co content, agreeing with the previous results for (Fe,Co)$_3$C alloys, see Fig.~\ref{fig:VCA_1D_MAE_ms_ml}.
The highest value of MAE (0.83~MJ\,m$^{-3}$; leading to magnetic hardness of 1.02) was determined for Co$_{3}$(C$_{0.3}$N$_{0.7}$).
However, alloying with N can further destabilize orthorhombic carbides.

%-------Co3(B,C,N) - MAE and magnetic moments----------
%
As we said, the most promising MAEs are observed for orthorhombic alloys with full Co content, see Fig.~\ref{fig:2D}.
To take a closer look at this range, in a one-dimensional  Fig.~\ref{fig:Co3(B,C,N)dep} we again show the dependence of MAE on element concentration at the 2$p$ position, along with a plot of spin magnetic moment.
At the center of B-C-N range is the ortho-Co$_3$C compound, known from previous discussion as a promising composition for hard magnetic applications. 
Although raising the nitrogen concentration leads to an increase in MAE, for reasons related to the instability of the ortho-Co$_3$N phase, a better approach may be to alloy ortho-Co$_3$C with boron, which should both stabilize the system and raise its Curie temperature.
Moreover, the dependence of the magnetic moment on the concentration of 2$p$ atoms is non-intuitive, see Fig.~\ref{fig:Co3(B,C,N)dep}(b).
While in the range from B to C it is qualitatively consistent with the ten percent decrease in magnetic moment measured for the isostructural Fe$_3$(B,C) phase~\cite{nicholson_solubility_1957}, in the range from C to N the spin magnetic moment shows a surprising minimum.
It comes from the minima of spin magnetic moments at both non-equivalent Co positions.
Light is shed on the minima in magnetic moments by density analysis, which in the case of the FPLO code involves determining the number of electrons in local density functions.
The excess charge on Co atom relative to the neutral cobalt atom is shown in Fig.~\ref{fig:Co3(B,C,N)dep}(c).
The electrons from the cobalt atoms transfer to the 2$p$ atoms.
The observed minimum in the magnetic moment correlates with the charge minimum.
Additional calculations without spin polarization (non-magnetic) show that the charge minimum is primary to the magnetic moment.
The theoretical minimum discussed here for the orthorhombic phase may not be experimentally verifiable, since experiments to date for Co$_3$N$_{1+\delta}$~\cite{lourenco_stability_2014} and (Fe,Co)$_3$N~\cite{wu_iron-doped_2021} indicate that Co$_3$N prefers the hexagonal structure.

Furthermore, the calculated MAE of Co$_3$B of less than 0.2~MJ\,m$^{-3}$ is much lower than the corresponding room-temperature experimental value of 0.65~MJ\,m$^{-3}$~\cite{pal_properties_2017}.
The discrepancy is due to, among other things, the dependence of MAE on temperature, as the calculations were performed at 0~K.

%------Curie temperature---------------
%
As we discussed earlier, the experimental Curie temperature values for ortho-Co$_3$C were determined for nanoparticles and range from 498~K to 650~K~\cite{mikhalev_magnetic_2019,harris_high_2010,turgut_metastable_2016,el-gendy_enhanced_2014} and 
are also significantly higher than the value for solid ortho-Fe$_3$C of 481~K~\cite{nicholson_solubility_1957}.
Mean-field theory calculations, which customarily overestimate the Curie temperature, give for ortho-Co$_3$C a value of 458~K, suggesting that we should actually expect an experimental $T_C$ about 300~K for ortho-Co$_3$C.
The experimental Curie temperatures of ortho-Co$_3$B
are between 710~K~\cite{zieschang_magnetic_2019} and 750~K~\cite{pal_properties_2017}, much higher than for the isostructural carbide.
In Fig.~\ref{fig:tc_co3bcn}, we show the results of Curie temperature calculations in the ortho-Co$_3$(B-C-N) concentration range.
In agreement with the experiment, we observe a decrease in Curie temperature with the transition from Co$_3$B to Co$_3$C.
However, for the transition from Co$_3$C to Co$_3$N, we observe a maximum in the Curie temperature.
A similar maximum in $T_C$ was observed experimentally for $\epsilon$-Fe-N system with increasing nitrogen concentration~\cite{wriedt_fe-n_1987}.
As we mentioned above, since Co$_3$N seems to prefer the hexagonal phase, the presented $T_C$ results for the ortho-Co$_3$(C-N) system will most probably remain only interesting theoretical considerations.

%-------section summary
%
In summary, the alloying at the 2$p$ position, as  discussed in this section, allows for the control of MAE, magnetic moment, and Curie temperature.
Alloying the considered carbides with boron leads to an increase in the Curie temperature by several tens of percent.
The increase in Curie temperature during alloying with boron correlates with an increase in magnetic moment up to about 10\%.
For the more stable orthorhombic phases, the highest magnetic hardnesses ($\sim0.9$) are found near the Co$_3$C composition, while remaining similarly high until about 30\% boron alloying.
Furthermore, the observation that MAE can depend on the total number of electrons in the system can prove useful in the designing of new compositions of magnetically hard materials.

\section{Summary and Conclusions}

In this study, we employed a first-principles approach to investigate the influence of composition on the magnetic properties of Fe$_3$C-type alloys.
The objective was to identify alloys that exhibit enhanced magnetic hardness.
This work has been divided into four parts.
In the first, we have presented results 
for Fe$_3$C and Co$_3$C compounds, 
in the second for (Fe,Co)$_3$C alloys, 
in the third for Fe$_3$C and Co$_3$C alloys with transition metals, and 
in the fourth for (Fe,Co)$_3$(B-C-N) alloys.
In each instance, both hexagonal and orthorhombic structures were analyzed.
%
%----- compounds -------
%
%---Vxc------- 
%
For Fe$_3$C and Co$_3$C compounds, we examined what effect the form of the exchange-correlation potential has on the MAE and magnetic moment.
%
%---FSM-------
%
By determining the dependence of the MAE on the fixed spin (magnetic) moment, we have preliminarily identified the range of MAE that the systems under consideration can adopt under varying temperatures or alloy compositions.
The outcome of the investigation was not particularly encouraging.
Of the four compounds considered, none proved to be magnetically hard.
The highest magnetic hardness of 0.91 was predicted for ortho-Co$_3$C.
%
%--- (Fe,Co)3C alloys ----
%
%-----VCA---------
%
In order to investigate the magnetic properties of (Fe,Co)$_3$C alloys, we employed the virtual crystal approximation to calculate changes in magnetic properties as a function of Co concentration. 
This analysis led to the identification of magnetically hard Co-rich hexa-(Fe$_{0.1}$Co$_{0.9}$)$_{3}$C phase.
%
%--- TM substitutions ----
%
Furthermore, we considered a series of hexa-Fe$_3$C, ortho-Fe$_3$C, and ortho-Co$_3$C alloys with transition metals at a concentration of 1/12 on Fe or Co site.
Calculations of the Curie temperature and magnetic hardness do not indicate that alloying with transition metals increases their potential for permanent magnet applications.
%
%----- (Fe,Co)3(B-C-N) alloys -----
%
Finally, we found that a considerable proportion of the ortho-Co$_3$(B-C-N) alloys are magnetically hard, of which boron substitution raises the Curie temperature and improves stability.

The most universal result of this work is the interpretation of the two-dimensional dependence of MAE on concentration at both 3$d$ and 2$p$ positions of (Fe-Co)$_3$(B-C-N) alloys.
The diagonal stripes visible in the MAE maps are interpreted as the dependence of MAE on the total number of electrons in the system, which is weakly responsive to the details of the occupancies at individual sites.
This type of dependence offers a novel perspective on the design of compositions of hard magnetic materials.

\vspace{5mm}

\section*{Acknowledgments}
We acknowledge the financial support from the National Science Center of Poland under grants
DEC-2019/35/O/ST5/02980 -- PRELUDIUM~BIS~1 (J.S.-A.) and
DEC-2021/41/B/ST5/02894 -- OPUS~21 (J.R.-G. and MW).
Part of the calculation was made in the Poznan Supercomputing and Networking Centre (PSNC/PCSS).
We thank P. Leśniak and D. Depcik for compiling the scientific software and managing the computer cluster at the Institute of Molecular Physics, Polish Academy of Sciences.
We thank Z.~Śniadecki and K.~Synoradzki for reading the manuscript and providing valuable comments.

\end{sloppypar}

\bibliography{bib,fe3c_MW,Fe3C_Co3C_exp}    

\end{document}